\documentclass[
 reprint,
superscriptaddress,
 amsmath,amssymb,
 aps,
pra,
raggedbottom]{revtex4-2}

\usepackage{graphicx}
\usepackage{float}

\usepackage{dcolumn}
\usepackage{bm}
\usepackage{braket}
\usepackage{hyperref}

\usepackage{siunitx}
\usepackage{physics}
\usepackage{comment}
\usepackage{ amssymb }
\usepackage[cmyk]{xcolor}
\usepackage{booktabs} 

\hypersetup{hidelinks}

\begin{document}
\preprint{APS/123-QED}


\title{Intrinsic Phononic Dressed States in a Nanomechanical System}

\renewcommand{\andname}{\ignorespaces}

\author{M. Yuksel}
\thanks{These authors contributed equally}
\affiliation{Department of Applied Physics, California Institute of Technology; Pasadena CA, 91125, USA.}

\author{M. P. Maksymowych}
\thanks{These authors contributed equally}
\affiliation{Department of Applied Physics and Ginzton Laboratory, Stanford University; Stanford CA, 94305, USA.}

\author{O. A. Hitchcock}
\affiliation{Department of Physics, Stanford University; Stanford CA, 94305, USA.}

\author{F. M. Mayor}
\affiliation{Department of Applied Physics and Ginzton Laboratory, Stanford University; Stanford CA, 94305, USA.}

\author{N. R. Lee}
\affiliation{Department of Applied Physics and Ginzton Laboratory, Stanford University; Stanford CA, 94305, USA.}

\author{M. I. Dykman}
\affiliation{Department of Physics and Astronomy, Michigan State University; East Lansing MI, 48824, USA.}

\author{A. H. Safavi-Naeini}
\affiliation{Department of Applied Physics and Ginzton Laboratory, Stanford University; Stanford CA, 94305, USA.}

\author{M. L. Roukes}
\email{roukes@caltech.edu}
\affiliation{Department of Applied Physics, California Institute of Technology; Pasadena CA, 91125, USA.}
\affiliation{Department of Physics and the Kavli Nanoscience Institute, California Institute of Technology; Pasadena CA, 91125, USA.}
\affiliation{Department of Bioengineering, California Institute of Technology; Pasadena CA, 91125, USA.}

\begin{abstract} 

Nanoelectromechanical systems (NEMS) provide a platform for probing the quantum nature of mechanical motion in mesoscopic systems. This nature manifests most profoundly when the device vibrations are nonlinear and, currently, achieving vibrational nonlinearity at the single-phonon level is an active area of pursuit in quantum information science. Despite much effort, however, this has remained elusive. Here, we report the first observation of intrinsic mesoscopic vibrational dressed states. The requisite nonlinearity results from strong resonant coupling between an eigenmode of our NEMS resonator and a single, two-level system (TLS) that is intrinsic to the device material. We control the TLS \textit{in situ} by varying mechanical strain, tuning it in and out of resonance with the NEMS mode. Varying the resonant drive and/or temperature allows controlled ascent of the nonequidistant energy ladder and reveals the energy multiplets of the hybridized system. Fluctuations of the TLS on and off resonance with the mode induces switching between dressed and bare states; this elucidates the complex quantum nature of TLS-like defects in mesoscopic systems. These quintessential quantum effects emerge directly from the \textit{intrinsic} material properties of mechanical systems – without need for complex, external quantum circuits. Our work provides long-sought insight into mesoscopic dynamics and offers a new direction to harness nanomechanics for quantum measurements.  

\end{abstract}

\date{\today}

\maketitle

Nanoelectromechanical systems (NEMS) have emerged as a major type of dynamical mesoscopic system~\cite{bachtold2022mesoscopic,ekinci2005nanoelectromechanical}. Beyond their fundamental appeal - given their compact size, long coherence time, and ability to sensitively detect force and motion - they provide a platform for various applications in quantum technology~\cite{schwab2005putting,maccabe2020nano, arrangoiz2019resolving,bozkurt2023quantum}. Of special interest is the realization of a nonequidistant energy spectrum at the level of few vibrational quanta; this can facilitate single-phonon coherent control and thereby open new avenues for quantum information processing and sensing~\cite{pistolesi2021proposal,samanta2023nonlinear}.
An efficient way of achieving such control is based on strongly coupling mechanical vibrations to a two-level system (TLS). In previous work, the TLS employed has typically been a qubit~\cite{o2010quantum,wollack2022quantum,satzinger2018quantum,yang2024mechanical,bozkurt2024mechanical}. The ensuing mode-TLS dynamics has been carefully studied across various fields, from photonic systems~\cite{haroche2006exploring} to Josephson-junction-based systems~\cite{simmonds2004decoherence, lisenfeld2015observation, lisenfeld2019electric, chen2024phonon, grabovskij2012strain} to superconducting microwave cavities~\cite{sarabi2016projected, kristen2024giant}. 

While such \textit{composite }hybrid mode-TLS platforms involving a mechanical system have been successfully demonstrated, achieving strong coupling of a mechanical mode to an \textit{intrinsic TLS} has not yet been achieved. This is despite significant effort~\cite{pedurand2024progress,tavakoli2022specific,ramos2013nonlinear}, as well as an extensive and growing body of indirect evidence showing the strong effect of intrinsic TLSs on the dynamics of NEMS vibrational modes~\cite{bachtold2022mesoscopic,maccabe2020nano,Maksymowych2025FreqNoise,kamppinen2022dimensional,cleland2024studying}. Therefore attaining strong intrinsic mode-TLS coupling and the associated strong intrinsic nonlinearity at the single-phonon level in NEMS is both interesting and important. Its realization would not only expand the platform of available quantum device architectures, but would also provide a new approach to acquiring fundamental insights into the dynamics of mesoscopic systems.

A promising pathway to overcoming this challenge lies in harnessing the intrinsic TLS defects found in solid-state materials~\cite{muller2019towards}. At low temperatures, material defects manifesting as tunneling states and modeled by TLSs contribute significantly to the spectrum of low-energy excitations and to the dynamical response of solids~\cite{phillips1972tunneling,anderson1972anomalous,phillips1987two}. The standard tunneling model (STM) predicts that a TLS arises from atoms or groups of atoms that can tunnel between two nearly degenerate configurations, modeled as a double-well potential~\cite{black1977spectral}. While amorphous materials host an abundant number of TLSs, similar tunneling states can also manifest from atomic-scale defects within crystalline materials, on surfaces, and at material interfaces~\cite{mihailovich1992low,esquinazi2013tunneling,gruenke2024surface,gruenke2025surface}. 

Here, we report the first observation of the intrinsic mesoscopic dressed states that result from the strong coupling between the mechanical vibrations and an individual TLS defect of a NEMS resonator. We employ lithium niobate (LN) NEMS resonators that are isolated from the environment by phononic crystals (PnC) ~\cite{arrangoiz2018coupling, arrangoiz2019resolving, wollack2021loss} and cooled to millikelvin temperatures in a dilution refrigerator. The PnC structure achieves low decay rates of phonons and TLSs, while using LN allows us to effectively induce a static crystal strain via its piezoelectric properties. As we show, this enables tuning the TLS frequency with exquisite control. Long-lived phonons confined to sub-wavelength volumes combined with the TLS tunability enable several key findings. (i) We measure mode splitting and reveal a double avoided-crossing pattern which, as we show, is a characteristic feature of strong resonant coupling of a NEMS mode to TLS defects. TLSs with different level spacings that are strongly coupled to the mode are also observed. Their energy spectrum remains stable, both over time and with significant thermal/power cycling of the NEMS. This allows us to reproducibly address individual TLSs by tuning them in and out of resonance with the mode. (ii) By varying the coherent drive power and temperature, we reveal the Jaynes-Cummings (JC) energy ladder in the TLS-NEMS system. Notably, the formation of phononic dressed states is inherited directly from the intrinsic TLS rather than from the interaction with an external system~\cite{yang2024mechanical}. The distribution over the dressed states exhibits high sensitivity to phonon occupancy, highlighting the coupled system's potential for quantum thermometry. (iii) We observe random telegraph signals (RTS) in the TLS-NEMS system response, arising from the TLS frequency switching on- and off-resonance with the mode, suggesting complex nature of the defects. 

These results provide a direct proof of the coupling of individual TLSs to a nanomechanical mode. The simplicity of the platform -- essentially a nanoengineered lithium niobate film with metallic electrodes -- stands in stark contrast to previous demonstrations of strong coupling and nonlinearity in quantum acoustics, which have relied on complex superconducting qubit circuits, optical cavities, or precisely engineered hybrid quantum devices~\cite{teufel2011circuit,yang2024mechanical,bozkurt2024mechanical}. That quantum states of motion emerge naturally from intrinsic material properties in such a minimal structure is striking, and highlights the fundamental nature of phonon-TLS interactions and suggests a path to quantum acoustic systems, which are significantly simpler, smaller and more powerful than previously thought.

\begin{figure*}
	\centering
	\includegraphics[width=0.9\textwidth]{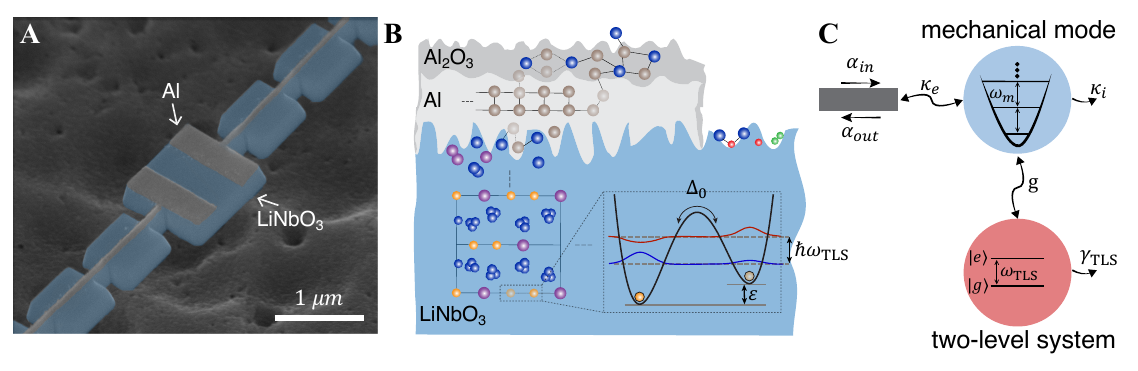} 

	\caption{\textbf{LiNbO\(_3\) nanomechanical resonator with TLS defects.}
	    (\textbf{A}) False colored scanning electron micrograph (SEM) of the lithium niobate (LiNbO\(_3\)) (blue) phononic crystal resonator with aluminum (gray) electrodes.
        (\textbf{B}) Toy model of the atomic structure in the cross section of the resonator. TLS can reside on the surface, within the bulk material or at the interfaces. Each TLS is modeled as a particle in a double-well potential with tunneling rate $\Delta_0$ and asymmetry energy $\varepsilon$. The two energy states are separated by the TLS frequency $\omega_\mathrm{TLS}$. (\textbf{C}) Schematic of the strong coupling model between the mechanical mode and an individual TLS defect, illustrating the interaction dynamics.}
	\label{fig1} 
\end{figure*}

\section*{The Strong coupling regime}

Our NEMS resonator is comprised of thin-film lithium niobate (LN, \SI{250}{\nm}) with aluminum (\SI{50}{\nm}) electrodes deposited on top (Fig.~\ref{fig1}A)~\cite{arrangoiz2018coupling}. The resonator is embedded in a suspended periodic array of cells, which form a PnC that supports a complete acoustic bandgap. The resonator is a ``defect'' in the PnC. Its acoustic eigenmode that we study is within the bandgap and is thus well-localized, which leads to a long mode lifetime. The aluminum electrodes facilitate piezoelectric transduction of mechanical vibrations~\cite{cleland2013foundations}. Here, we experiment with a $\sim 1.1\times\SI{1.1}{\micro\meter\squared}$ ``defect'' resonator with a fundamental shear mode frequency of $ \omega_m /2 \pi \approx \SI{1.873}{\GHz}$.

\begin{figure*}
	\centering
	\includegraphics[width=1\textwidth]{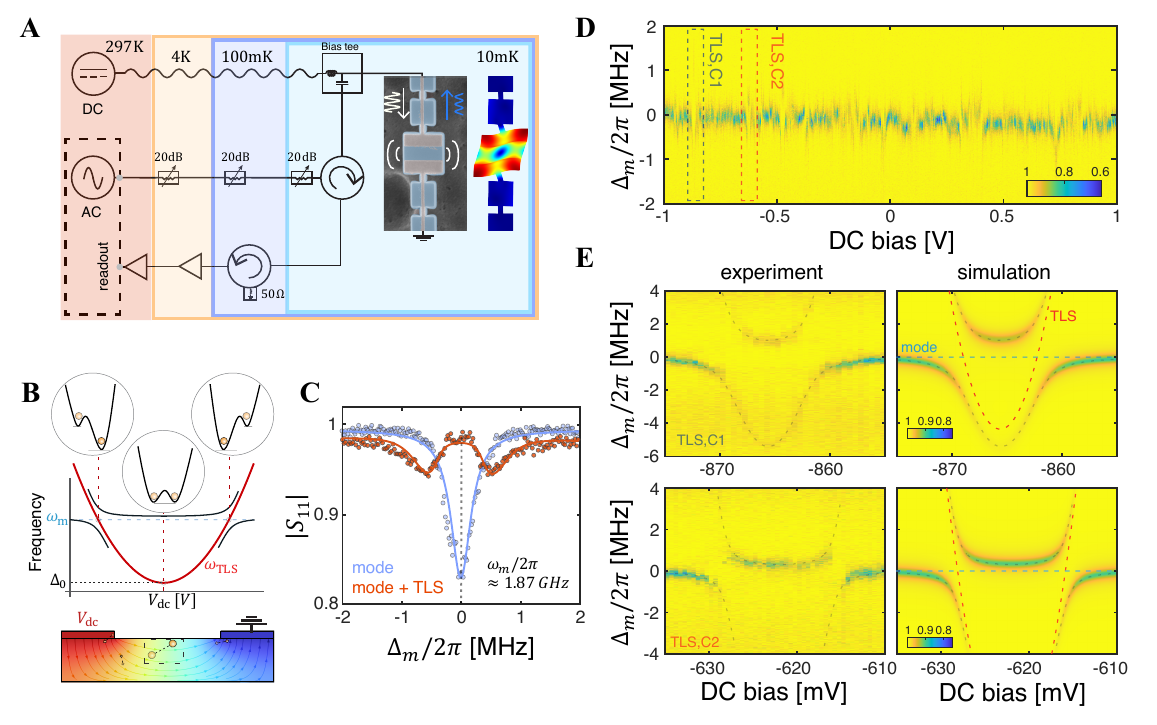} 
	
	\caption{\textbf{Strong coupling to individual TLS.}
        (\textbf{A}) Schematic of the microwave reflection measurement $S_{11}$ setup used to probe the fundamental shear mode localized at the center of the LN PnC resonator in our dilution refrigerator. 
	    (\textbf{B}) Tuning mechanism of the TLS frequency ($\omega_\mathrm{TLS}$) via the DC bias $V_\text{dc}$. The anticipated linear dependency of $\varepsilon$ on the static electric field results in a hyperbolic dependency of TLS frequency on DC bias,  $\omega_\mathrm{TLS} (V_\text{dc})$. Consequently, two distinct $V_\text{dc}$ values yield the same $\omega_\mathrm{TLS}$, corresponding to mirrored configurations of the double-well potential about $\Delta_0$. 
        (\textbf{C}) Example $\abs{S_{11}}$ spectra for large  (blue) and zero (orange) detuning between the NEMS and a strongly coupled TLS, characterized by a bare Lorentzian resonance and Rabi splitting, respectively. 
        (\textbf{D}) The ${|S_{11}|}$ spectra near the mode frequency ($\Delta_m = \omega_m - \omega_d$) as a function of $V_\text{dc}$ reveal many TLS crossings at different bias voltages.
        (\textbf{E}) Fine $V_\text{dc}$ sweeps near the two double avoided crossings -- labeled TLS,C1 (top) and TLS,C2 (bottom)-- associated with coupling rates $\mathrm{g}/2\pi$ of $2.4$ and $1.7~\text{MHz}$, respectively. Experimental results (left) are reproduced with simulations (right). Minima of the simulated $\abs{S_{11}}$ are overlaid on the experimental data as black dashed lines. The TLS and the mechanical mode frequencies are indicated on the simulated spectra as red and blue dashed lines, respectively.}
	\label{fig2} 
\end{figure*}

The structurally disordered aluminum-LN interface, amorphous LN leftover from fabrication, aluminum oxide, and the LN crystal itself can host numerous TLS defects, which could couple strongly to the local acoustic field. A conventional TLS model is sketched in Fig.~\ref{fig1}B as a particle in a double-well potential with asymmetry energy $\varepsilon$ and a tunneling energy $\Delta_0$ between the two minima. The ground $\ket{g}$ and excited $\ket{e}$ eigenstates are superpositions of the intra-well states with an energy splitting $\hbar \omega_{\text{TLS}} = \sqrt {\Delta_0^2 + \varepsilon^2}$, where $\omega_\mathrm{TLS}$ is the TLS transition frequency. 

\setlength{\abovedisplayskip}{2pt}
Previous studies on PnC resonators have shown anomalous frequency red-shifts and increased losses below $\sim \SI{1}{\kelvin}$~\cite{wollack2021loss,cleland2024studying, emser2024thin}, arising from coupling to a quasi-continuum of TLS defects with a broad distribution in frequency. In contrast, when an individual TLS is strongly coupled with the mechanical mode, the system's dynamics become dominated by this localized interaction rather than ensemble effects. An oscillatory strain field resonant with $\omega_{\text{TLS}}$ can induce transitions from $\ket{g}\to\ket{e}$, allowing energy exchange at a coupling rate $\mathrm{g}$. The resonant mechanical mode-TLS interaction is depicted in Fig.~\ref{fig1}C, where $\kappa_i$ is the intrinsic mechanical loss rate, $\kappa_e$ is the external coupling rate to the microwave waveguide and $\gamma_\text{TLS}$ is the TLS relaxation rate. This coupled system is described by the JC Hamiltonian~\cite{jaynes1963comparison}. In the frame of a resonant drive at frequency $\omega_d$, assuming the rotating wave approximation, this Hamiltonian reads:

\begin{align}
    \frac{H}{\hbar} = & \, \Delta_\text{m} a^\dagger a + \Delta_{\text{TLS}} \sigma_+ \sigma_- + \mathrm{g} \left( a \sigma_+ + a^\dagger \sigma_- \right) \notag \\
    & + i \sqrt{\kappa_e}  \alpha_{\text{in}} \left( a - a^\dagger \right)
    \label{eq:Hamiltonian}
\end{align}
\noindent 
where \(\Delta_m=\omega_m-\omega_d\) and $\Delta_{\text{TLS}}=\omega_{\text{TLS}}-\omega_d$ are the detunings, $\sigma_+$ ($\sigma_-$) are the TLS raising (lowering) operators, $a$ ($a^\dag$) are the phonon annihilation (creation) operators, and $\alpha_\text{in}\approx \sqrt{P/\hbar\omega_d}$ is the amplitude of the drive field related to the coherent drive power $P$ incident to the mode. 

To explore individual TLSs, we thermalize the sample chip to our dilution refrigerator at $T \approx \SI{10}{\milli\kelvin}$ ($k_B T / \hbar \ll \{ \omega_m, \omega_\mathrm{TLS} \}$). A simplified schematic of our microwave reflection measurement setup is shown in Fig.~\ref{fig2}A. We apply a coherent microwave drive at low power, $P = -150 \text{ dBm}$, ensuring the intracavity phonon number is near zero. By sweeping $\omega_d$ across the mode frequency $\omega_m$, we measure the reflection coefficient ($S_{11}$) and capture the steady-state response of the resonator. We observe a decrease in reflected power near-resonance, corresponding to energy absorption by the mechanical mode.

To reconfigure the TLS bath and bring an individual TLS into resonance with the mechanical mode, we apply a DC bias voltage $V_\text{dc}$ across the aluminum electrodes of our device (Fig.~\ref{fig2}B). In the piezoelectric LN, this creates both a static electric field $\vec{E}$ and an associated strain field $\vec{S} = \hat d \vec{E}$, where $\hat d$ is the piezoelectric tensor~\cite{cleland2013foundations}. These fields perturb the local TLS environment, modifying the well asymmetry $\varepsilon$. To first order, the response to the applied bias is linear~\cite{grabovskij2012strain, lisenfeld2015observation}, resulting in a DC-dependent TLS frequency of:

\begin{align}
    \omega_{\text{TLS}} = \frac{1}{\hbar}\sqrt{\Delta_0^2 + \left(\varepsilon_0 + \frac{\partial \varepsilon}{\partial V_{\text{dc}}} V_{\text{dc}} \right)^2}\
	\label{eq:TLS} 
\end{align}

\noindent where $\partial \varepsilon / \partial V_\text{dc}$ is the TLS sensitivity to DC voltage and $\varepsilon_0$ is the intrinsic asymmetry energy in the absence of the applied DC field. As seen from Fig.~\ref{fig2}B and the symmetry of Eq.~\ref{eq:TLS}, there are two distinct values of $V_\text{dc}$ that result in the same $\omega_\mathrm{TLS}$. Similar methods have achieved in-situ control of TLS frequencies in superconducting qubits and microwave resonators, \textit{e.g.} by bending the device chip~\cite{bilmes2022probing,meissner2018probing,kristen2024giant,grabovskij2012strain} or via DC-electric field biasing~\cite{lisenfeld2019electric}. 

When all strongly coupled TLSs are far detuned from the mode, we observe a linear resonator response as shown in Fig.~\ref{fig2}C (blue). When a strongly coupled TLS is brought into resonance with the mechanical mode, we observe Rabi splitting (Fig.~\ref{fig2}C, orange). In Fig.~\ref{fig2}D, we display $|S_{11}|$ as a function of drive detuning ($\Delta_m$) over a wide range of $V_\text{dc}$. We observe multiple avoided crossings at different $V_\text{dc}$, corresponding to different TLS defects tuned into resonance with the mechanical mode. Fig.~\ref{fig2}E presents measurements with finer DC steps over a narrower range, revealing clear evidence of two double avoided crossings each associated with an individual TLS, denoted TLS,C1 and TLS,C2 centered at $\sim\SI{-865}{\milli\volt} \text{ and } \sim\SI{-623}{\milli\volt}$, with mode-TLS coupling rates $\mathrm{g}/2\pi$ of $ \SI{2.4}{\MHz} \text{ and } \SI{1.7}{\MHz}$, respectively. 

To quantitatively analyze our observations, we model the coupled TLS-phonon system including environmental interactions. An analytical expression for $S_{11}$ can be derived in the low-temperature regime ($n_\text{th} =0$, where $n_\mathrm{th} = [\exp(\hbar\omega_m/k_BT)-1]^{-1}$) 
~\cite{sarabi2015cavity, kristen2024giant}. A formulation for arbitrary $n_\mathrm{th}$ was given in ~\cite{dykman1976spectral}. Here, we provide numerical results equivalent to this formulation. The master equation for the density matrix $\rho$ of the coupled TLS-mode system is~\cite{blais2021circuit}:

\begin{align}
    \dot{\rho} = & -\frac{i}{\hbar} [H,\rho] 
    + (n_{\text{th}}+1)\kappa_i D[a]\rho 
    + n_{\text{th}} \kappa_i D[a^\dagger]\rho \notag \\
    & + \gamma_\text{TLS}(n_{\text{th}}+1)D[\sigma_-]\rho 
    + \gamma_\text{TLS} n_{\text{th}} D[\sigma_+]\rho
    \label{eq:EOM}
\end{align}

\noindent 
where $D[x]$ is the Lindblad damping superoperator defined by $D[x]\rho = x\rho x^\dagger - \frac{1}{2} \left( x^\dagger x \rho + \rho x^\dagger x \right)$ and $H$ is given by Eq.~\ref{eq:Hamiltonian} with Eq.~\ref{eq:TLS} used for $\omega_\mathrm{TLS}$. We calculate the steady-state expectation value of the mode annihilation operator: $\langle a \rangle_{\text{ss}} = \text{Tr}(\rho_{\text{ss}} a)$ via QuTiP ~\cite{johansson2012qutip}. For a device weakly and linearly coupled to a waveguide, the output field can be written as  $\alpha_{\text{out}} = \alpha_{\text{in}} + \sqrt{\kappa_e} \langle a \rangle_{\text{ss}}$ ~\cite{gardiner1985input}. We then calculate the simulated reflection coefficient by $S_{11} = \frac{\alpha_{\text{out}}}{\alpha_{\text{in}}}$.

The simulated reflection spectra (see parameters in Supplementary Table \ref{tab:ps_fits}) accurately reproduce the observed avoided crossings in Fig.~\ref{fig2}E, validating the presence of strong coupling. In particular, the measurement featuring TLS,C1 clearly demonstrates the hyperbolic shape of the TLS trajectory ($\omega_\text{TLS,C1}(V_\text{dc})$), which aligns with the theoretical model in Eq.~\ref{eq:TLS} (Fig.~\ref{fig2}B). This corroborates the linear dependence of $\varepsilon$ on $V_\text{dc}$ and the double-well potential model for a TLS coupled to a mechanical mode. The decay rates of the mode and TLSs are determined from the best-fit estimates of the spectra, with values of $\kappa_i/2\pi \approx 400 \text{ kHz and } \gamma_\text{TLS}/2\pi \approx 200 \text{ kHz} $ for TLS,C1. We estimate single-phonon cooperativities ($C=4\mathrm{g}^2/\kappa\gamma_{\text{TLS}}$, where $\kappa = \kappa_i+\kappa_e$) of $\sim 3500$ for TLS,C1 and  $\sim 1800$ for TLS,C2 (see \ref{sec:StrongCouplingSims} for details). 

We measured strong coupling to individual TLSs in multiple LN resonators of varying dimensions and using different experimental setups located at both Caltech and Stanford (see \ref{sec:StanfordMeasurements}). This highlights the robustness of our device architecture for studying and exploiting strong phonon-TLS interactions for quantum experiments.

\section*{The Jaynes-Cummings Energy Levels}     

\begin{figure*} 
	\centering
	\includegraphics[width=0.9\textwidth]{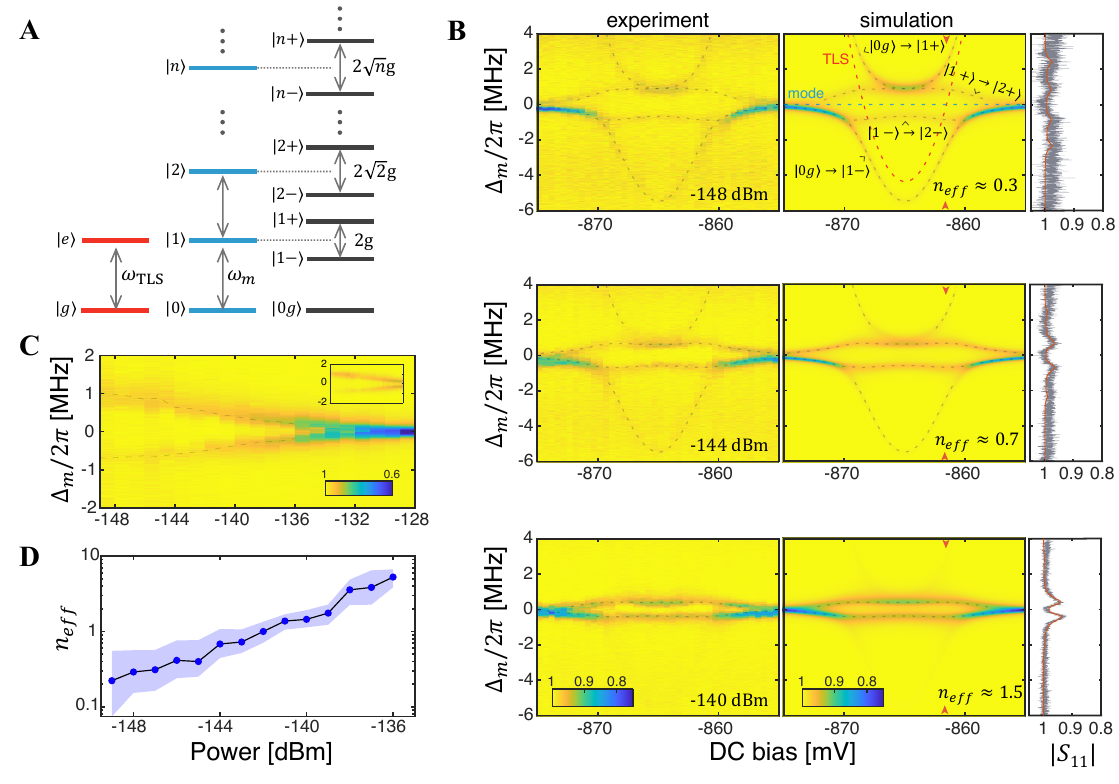} 
	\caption{\textbf{Nonlinear response.}
	    (\textbf{A}) Jaynes-Cummings energy levels for the on-resonance case ($\omega_m = \omega_\text{TLS}$) of the strongly coupled TLS-mechanical mode system. 
        (\textbf{B}) Experimental and simulated spectra of $|S_{11}|$ for TLS,C1 at three different input powers at the base temperature (10~mK). The additional fine structures appearing in the spectrum corresponds to the higher order transitions, revealing the TLS induced nonlinearity. Dashed lines on the experimental spectra denote minima extracted from the simulated $|S_{11}|$. On the far right, $|S_{11}|$ data are shown for detunings near resonance ($\omega_m \approx \omega_{TLS}$) - highlighted by red arrows in the simulated spectra - and the simulated response (orange) is overlaid on the experimental data (gray). As the incident microwave power increases (top to bottom), the response transitions from characteristic \textit{avoided crossing} to an \textit{eye-like} crossing.
        (\textbf{C}) The spectra showing the transition from a frequency-split response to a single resonance response under increased input power acquired at fixed $V_\text{dc}$. Dashed lines are the minima of the simulated $|S_{11}|$ spectra shown in the inset.
        (\textbf{D}) Plot of $n_{eff}$ versus input drive power. The shaded region encompasses $n_{eff}$ values that produce simulation results associated with a least-squares error within $5\%$ of the lowest error.}
	\label{fig3} 
\end{figure*}

While strong coupling to an individual defect is evident, the quantum nature of these defects described by the JC model has not been fully established. In this section, we study the response of the strongly coupled TLS-NEMS device at higher excitation numbers. Under strong coupling, such excitations occupy dressed states rather than simple bare states. In the on-resonance case ($\omega_m = \omega_\text{TLS}$), each pair of dressed states in level $n$ (for $n \geq 1$) is formed by the symmetric $\ket{n+} \equiv ( \ket{n,g}+\ket{n-1,e})/\sqrt{2}$ and antisymmetric $\ket{n-} \equiv ( \ket{n,g}-\ket{n-1,e})/\sqrt{2}$ combinations, where $n$ enumerates the vibrational quanta of the mode. In a more general case ($\omega_m \neq \omega_\text{TLS}$), the dressed state energies are $\hbar[\omega_m n + \frac{1}{2}\delta \pm (\mathrm{g}^2 n+ \delta^2/4)^{1/2}]$, where $\delta = \omega_\mathrm{TLS} - \omega_m$ ~\cite{dykman1976spectral}. The frequency splitting between dressed states scales as  $2 \sqrt{n}\mathrm{g}$ (Fig.~\ref{fig3}A), a characteristic of the JC model in the strong coupling regime. 

These dressed states are well-known in quantum optics ~\cite{brune1996quantum}, and have been explored using microwave superconducting qubits ~\cite{schuster2007resolving,bishop2009nonlinear,fink2008climbing}. In the previous section, we demonstrated Rabi splitting at $n = 1$, corresponding to the $\ket{0g} \to \ket{1-}$ and $\ket{0g} \to \ket{1+}$ transitions. Now, in our phononic TLS-NEMS platform, we provide direct evidence of the Rabi splitting for the transitions between the first and second excited multiplets of the dressed states. We also show that, as the excitation number increases, the response changes to the expected classical behavior~\cite{ginossar2012nonlinear,peano2010quasienergy}.

Our studies of the excited dressed states are twofold: we $1)$ vary the input drive power $P$ at the base temperature ($\SI{10}{\milli\kelvin}$), and $2)$ controllably sweep the temperature at a fixed $P$. At each power and temperature, we repeat the TLS crossing measurements, then model the results using simulations with Eq.~\ref{eq:TLS} and Eq.~\ref{eq:EOM}, as we did earlier.

Figure~\ref{fig3}B illustrates how the spectral features of the TLS,C1 crossing evolve at three different powers. At $P= -148\text{ dBm}$, in addition to a doublet response at the avoided crossings, the appearance of quadruplets of lines is a hallmark of higher energy transitions in JC ladder ($n>1$). New frequencies in the spectrum correspond to the transitions $\ket{1-}\to \ket{2-}$ and $\ket{1+}\to \ket{2+}$. The corresponding spectral lines are marked in the simulated spectra by black dashed lines. 

As the input power increases, so does also the contribution from transitions between higher energy levels, because the transition amplitude increases with the level number. The overall spectrum shifts toward the corresponding transition frequencies. However, the widths of the spectral lines corresponding to individual transitions increases with the level number and ultimately the transitions become unidentifiable~\cite{dykman1976spectral}. Consequently, the conventional \textit{avoided crossing} pattern gradually transitions to an \textit{eye-like} spectrum as the higher energy level transitions dominate, a phenomenon also observed in spin-circuit quantum electrodynamics (QED)~\cite{bonsen2023probing}. Eventually, the spectral lines merge together and the system's behavior becomes classical ~\cite{peano2010quasienergy,ginossar2012nonlinear}. In our system, this transformation of the spectrum is seen in Fig.~\ref{fig3}~B,C for increasing values of the input power.  

Associated with the growing population of higher energy levels due to resonant pumping is an overall heating of the resonator. The experimental observations can be reproduced by treating the bath temperature as a power-dependent parameter. Thus, we replace $n_\mathrm{th}$ on our model with an effective phonon number $n_{eff}$, while also adjusting $\kappa_i, \gamma_{TLS}, \kappa_e$ in the master equation and estimate these parameters by iteratively minimizing the error between simulated and experimental spectra (see \ref{sec:StrongCouplingSims}). Fig.~\ref{fig3}B displays the simulated spectra alongside the experimental data, demonstrating excellent agreement with the corresponding $n_{eff}$ values indicated in the plots. Movie S1 further illustrates the evolution of the TLS,C1 crossing as the input power is increased from $-150~\text{dBm}$ to $-135~\text{dBm}$ in $1~\text{dB}$ increments.  Fig.~\ref{fig3}C demonstrates spectrum of $|S_{11}|$ near $\omega_m$ as a function of power at a fixed DC ($V_\text{dc} = \SI{-860}{\milli\volt}$), corresponding to the near-resonance condition ($\omega_m \approx\omega_\text{TLS}$). A steady decrease in the frequency splitting with increasing $P$ marks the transition to a classical response. Simulated results, shown in the inset Fig.~\ref{fig3}C, closely align with the experimental data. Fig.~\ref{fig3}D plots the $n_{eff}$ used in the simulations for each power. 

 \begin{figure*}
	\centering
	\includegraphics[width=1\textwidth]{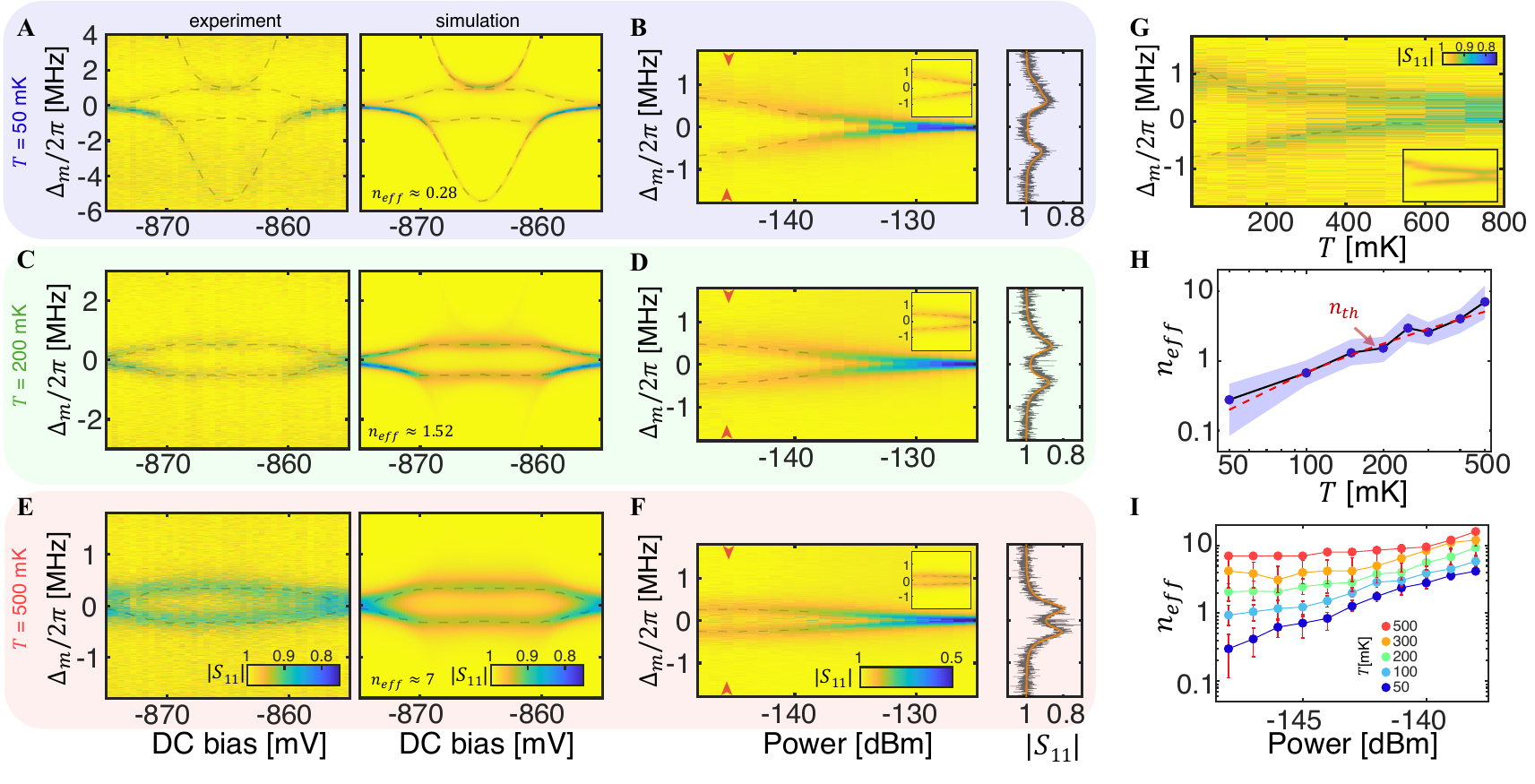} 
	
	\caption{\textbf{Temperature Effect.}
	  (\textbf{A}) $\abs{S_{11}}$ measurements of TLS,C1 at low input power $P = -150\text{ dBm}$ and at $T = 50~\text{mK}$. The dashed lines are the $|S_{11}|$ minima at each DC bias obtained from simulations.
      (\textbf{B}) Power dependent $|S_{11}|$ response acquired with the fixed DC bias that puts $\omega_m \approx \omega_{TLS}$,  demonstrating the transition to a classical response. The inset shows the corresponding simulation results, the minima of which are overlaid as black dashed lines on the measured spectra. In the right panel, the $\abs{S_{11}}$ spectrum for $P = -145\text{ dBm}$ (marked by red arrows) is plotted with the corresponding fit (orange). Panels 
      (\textbf{C}, \textbf{D})  and (\textbf{E}, \textbf{F}) show the same measurements as (\textbf{A}, \textbf{B}), acquired at 200~mK and 500~mK, respectively.
      (\textbf{G}) $|S_{11}|$ spectra acquired with a fixed DC bias ($\omega_m \approx \omega_{TLS}$) and low power ($P = -150~\text{dBm}$) as temperature increases. The dashed line is a fit from the simulations displayed in the inset. The $n_{eff}$ used in the simulations is shown in (\textbf{H}), which overlaps with the Bose-Einstein predicted thermal occupancy ($n_{\text{th}}$), highlighting good thermalization of the device. The blue area indicates the $5\%$ uncertainty range. (\textbf{I}) The $n_{eff}$ values used in simulations to reproduce the experimental data at varying power and temperature.}
	\label{fig4} 
\end{figure*}

In the next set of experiments, we apply a weak drive power ($P=-150 \text{ dBm}$) while controllably increasing the stage temperature. Fig.~\ref{fig4}A, C, and E display the measured and simulated TLS,C1 crossings for the temperatures of $\SI{50}{\milli\kelvin}$, $\SI{200}{\milli\kelvin}$, and $\SI{500}{\milli\kelvin}$, respectively. As the temperature increases, we observe the emergence of a quadruplet of spectral lines and a gradual transition to the classical response at more elevated temperatures. This behavior closely mirrors that reported in cavity QED systems~\cite{fink2010quantum,rau2004cavity}, and aligns with the power sweep measurements demonstrated earlier. Figure~\ref{fig4}G shows the $|S_{11}|$ spectrum as a function of temperature for a fixed $V_\text{dc}$, with simulated results in the inset. The estimated $n_{eff}(T)$ closely follows the expected thermal phonon occupation $n_{\text{th}}(T)$ (Fig.~\ref{fig4}H) when we take the stage temperature as a reference. This indicates that the hybrid system remains well-thermalized with the stage under weak driving conditions. In Fig. \ref{fig4}B, D, and F, we examine the gradual transition from the split to the classical response on-resonance ($\omega_m \approx \omega_{\text{TLS}}$) as we sweep $P$ at the different stage temperatures. On the right side of each panel, we provide a specific $|S_{11}|$ trace at $P = -145\text{ dBm}$. Figure \ref{fig4}I plots the $n_{eff}$ used to simulate the experimental data for the power sweeps which suggests the coupled system is more sensitive to changes in coherent drive power at low temperatures, whereas at higher temperatures the spectrum shows constant splitting up to certain power before it gradually decreases (Fig.~\ref{fig4}F). The high sensitivity of the spectrum to temperature suggests a potential application in thermometry. 

\section*{State Switches of a Mechanical TLS}

\begin{figure*} 
	\centering
	\includegraphics[width=0.95\textwidth]{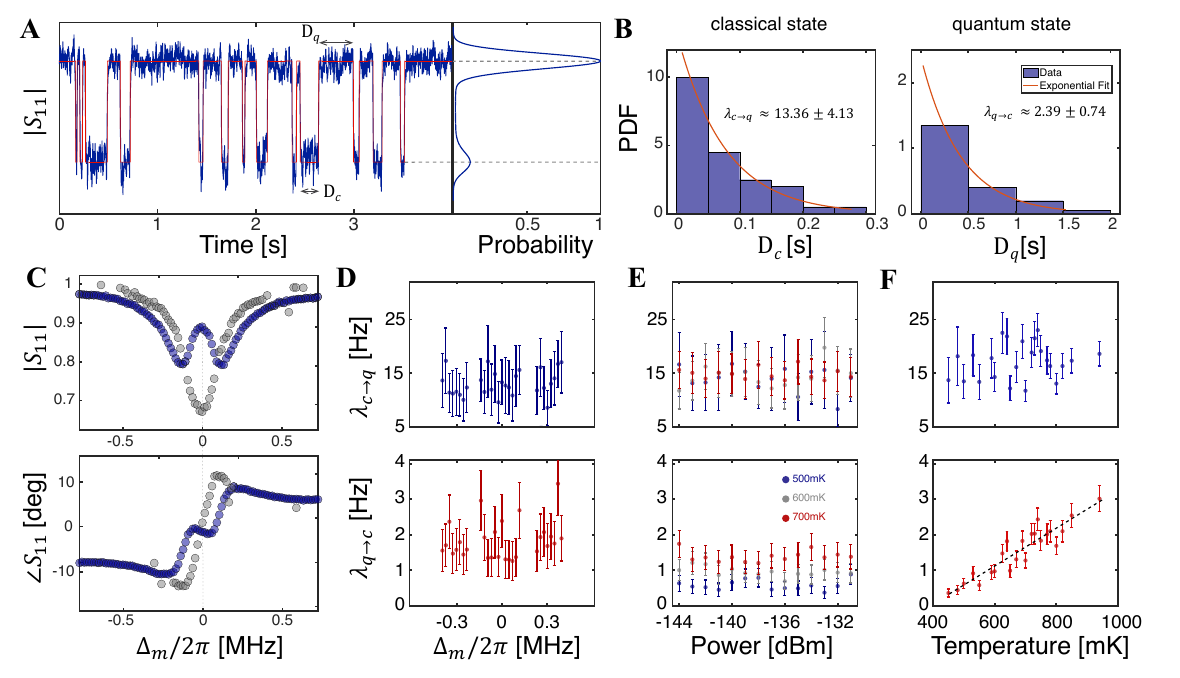} 
	\caption{\textbf{Random Telegraph Signal.}
	    \textbf{(A)} Time series data of $\abs{S_{11}}$ recorded at $600\text{ mK}$ with a fixed drive frequency ($\omega_d \approx \omega_m$) reveals clear fluctuations between two stable states. In the right panel, the probability density function (normalized to its maximum value) of the time series data is shown. The two distinct peaks in the density are used to generate the ladder plot (red) superimposed on the raw time series data (blue), highlighting the switching events. 
	    \textbf{(B)} Histogram of the time durations spent in the lower ($\mathrm{D}_{c}$) and higher amplitude ($\mathrm{D}_{q}$) states, extracted from the time series data. An exponential fit (red curve) is applied to determine the switching rates between the states, $\lambda_{c \to q}$ and $\lambda_{q \to c}$. 
	    \textbf{(C)} $|S_{11}|$ and $\angle S_{11}$ are constructed using the peak values from the probability density of each time series data across different $\Delta_m$. This reveals that the observed fluctuations correspond to the system transitioning between quantum and classical responses.
	    \textbf{(D)} Switching rates as a function of detuning around the mechanical resonance show no significant dependence.
        \textbf{(E)} Dependence of switching rates on input power for three different stage temperatures. 
        \textbf{(F)} Temperature dependence of the switching rates.}	
	\label{fig5} 
\end{figure*}

Along with the frequency splitting discussed earlier,we observe switching of the $S_{11}$ spectrum between a split response and a classical linear resonator response. These fluctuations occur across a range of DC tunings and temperatures and are observed with various resonators (see \ref{sec:additional_data}). The fluctuations we study here are observed with TLS,C1 when we increase the temperature. 

We characterize these slow fluctuations through time-domain reflection measurements, $S_{11} (t)$, taken at a fixed drive frequency $\omega_d$ close to $ \omega_m\approx\omega_\mathrm{TLS}$. Figure~\ref{fig5}A demonstrates RTS observed at $T = 600\text{ mK}$, $P = - 135\text{ dBm}$, and $\Delta_m = 0$. First, we extract the probability density function (PDF) of $\abs{ S_{11} (t)}$ using kernel density estimation (KDE). The presence of RTS yields two distinct peaks. From these peaks, we construct a ladder plot by assigning each data point to its nearest peak value, and then overlay it on the experimental data in Fig.~\ref{fig5}A (see red line). We then identify the switch instants and calculate the time spent (dwell-time) in the split spectrum (the quantum response) and in the classical single-lorentzian response, denoted as $\mathrm{D}_q$ and $\mathrm{D}_c$, respectively. By analyzing numerous switching events, we gather sufficient statistics to study their probability distribution. In Fig.~\ref{fig5}B, we estimate the PDF of the dwell times by constructing histograms, which fit well to an exponential distribution of the form ($\lambda e^{-\lambda t}$). This indicates that the switching process follows Poisson statistics with a rate $\lambda$. From these fits, we extract the switching rates $\lambda_{q \to c}$ and $\lambda_{c \to q}$, corresponding to the transitions from quantum to classical and classical to quantum responses, respectively. We can now examine how these rates and jump amplitudes vary with different experimental parameters.

We collect time-domain traces of $S_{11} (t)$ at different drive frequencies ($\omega_d$) across $\omega_m$. For each frequency, we obtain a probability distribution of $S_{11}(t)$ and extract the values at the peaks. In Fig.~\ref{fig5}C, each data point represents the peak value obtained at a given detuning $\Delta_m$ (see \ref{SI_fig:RTS_TLSC1}), with the higher probability peak shown in blue and the lower in gray. Strikingly, the resulting spectra reveal two stable values of $|S_{11}|$ and the phase shift of the reflected signal $\angle S_{11}$, corresponding to the classical linear resonator response (gray) and the split response (blue) of the NEMS-TLS system. This switching behavior remains highly stable over a wide temperature range (see \ref{SI_fig:bistability_temperature}). We do not observe RTS away from the mode's frequency. Unlike the switch amplitudes,  Fig.~\ref{fig5}D indicates that the switching rates show no significant dependence on frequency detuning. Moreover, $\lambda_{c \to q} > \lambda_{q \to c}$ indicates the system response is predominantly the split response.

In Fig.~\ref{fig5}E, we fix $\omega_d$ on resonance ($\Delta_m=0$) and vary the coherent drive power $P$ within the range where split-response remains observable. The switching rates obtained at three different stage temperatures remain insensitive to power. We note here that since the thermal bath is already at high temperatures, the power effect  is minimal for the considered power range, as we demonstrated in the previous section. In Fig.~\ref{fig5}F, we collect the RTS data at a fixed power while increasing the stage temperature and find that $\lambda_{q \to c}$ increases linearly with temperature, reflecting a higher probability of observing the classical response, while $\lambda_{c \to q}$ remains largely unaffected. This suggests that the $q\to c$ transitions are thermally initiated, while there is a faster and weakly dependent on temperature relaxation process that leads to $c\to q$ transitions.
 
The RTS can be explained by the TLS frequency fluctuating between on- and off- resonance with the mechanical mode, resulting in a split spectrum (on-resonance) or a classical response (off-resonance). Similar telegraphic switching has been reported in superconducting qubits and microwave devices, where it was attributed to coupling of TLS to an off-resonant two-level fluctuator (TLF), which undergoes thermally driven random transitions between its eigenstates ~\cite{burnett2014evidence,bejanin2022fluctuation,meissner2018probing,de2021quantifying,faoro2015interacting}. While a similar mechanism may be involved in our system, as the switching shows sensitivity to temperature but not to resonant drive power, the complex dynamics of TLSs at elevated temperatures and the presence of multiple TLSs make it difficult to definitively identify a single fluctuator. Furthermore, we demonstrate in \ref{SI_fig:RTS_R5} direct control over the switch rate statistics and show that RTS noise can be removed entirely through DC biasing. Our NEMS platform, which exhibits remarkably stable decoherence processes, enables future studies on understanding and mitigating noise. For instance, in a similar NEMS device~\cite{Maksymowych2025FreqNoise}, we study RTS caused by a single far off-resonant TLS. 

\section*{Conclusion}

The observation of strong coupling in a NEMS resonator with atomic-scale defects marks the first successful observation of phonon-TLS coupling in a solely \textit{mechanical} system. Our findings provide new insights into TLS-phonon interactions, particularly the nature of mechanical decoherence, thermalization and drive-induced heating in the single-phonon regime where quantum applications are realized. Our ability to control and populate higher energy states of the hybridized system could enable the preparation and control of more exotic quantum states stored purely within mechanical vibrations. Furthermore, the TLS-induced nonlinear scaling of the energy levels can be engineered as a highly sensitive quantum thermometer.

The robustness, reproducibility, and stability of our platform is validated through repeated measurements across different setups and resonators at both Caltech and Stanford. The ability to access and manipulate individual TLS defects within mechanics opens the possibility of using them as mechanical qubits for quantum computing and for advances in quantum memories, transducers, and quantum sensing applications. The observation of these quintessentially quantum effects in such a minimal platform -- requiring only a nanostructured piece of lithium niobate -- demonstrates that quantum acoustics can arise naturally from the intrinsic couplings within a material, without the need for superconducting or complex electromagnetic circuitry or hybrid quantum integration. Combined with the robustness and reproducibility of our observations, the simplicity of our setup suggests that other approaches to quantum-engineered acoustic fields for sensing and computing may be worthwhile.

\section*{Acknowledgements}

We are grateful for support for this work from the Gordon and Betty Moore Foundation, Grant 12214 (M.L.R., A.H.S.-N, \& M.I.D.) and a Moore Inventor Fellowship (A.H.S.-N.); the Wellcome Leap – Delta Tissue Program (M.L.R.); the National Science Foundation, Award 2016555 (M.L.R.) and CAREER award ECCS-1941826 (A.H.S.-N.); Defense Advanced Research Projects Agency (DARPA) under cooperative agreement HR0011-23-2-0004 (M.I.D); Amazon Web Services, Inc. (A.H.S.-N.); and the US Department of Energy, Grant DE-AC02-76SF00515 (A.H.S.-N.). M.P.M. is grateful for support from an NSERC graduate fellowship. The authors thank the Stanford Nanofabrication Facility (supported by National Science Foundation awards ECCS-2026822 and ECCS-1542152), Caltech’s Kavli Nanoscience Institute, and Stanford’s Q-NEXT Center. We also acknowledge valuable discussions with W. Fon, S. Habermehl, and A. Nunn, and fabrication assistance from W. Jiang, K.K.S. Multani and A.Y. Cleland.

\section*{Competing interests}
A.H.S.-N. is an Amazon Scholar. The other authors declare no competing interests.

\clearpage 
\bibliographystyle{naturemag}
\bibliography{main}\newpage
\clearpage
\newpage
\onecolumngrid

\section*{Supplementary information}

\renewcommand{\figurename}{}
\renewcommand{\thesection}{Supplementary Note \arabic{section}}
\setcounter{section}{0}
\renewcommand{\thefigure}{Supplementary Figure \arabic{figure}}
\setcounter{figure}{0}

\renewcommand{\tablename}{Supplementary Table}
\setcounter{table}{0}

\renewcommand \theequation{S\arabic{equation}}
\setcounter{equation}{0}

\section{Device Details}

Our devices are fabricated from a 250 nm thick X-cut congruent lithium niobate (LN) wafer bonded to a high resistivity $\langle 111 \rangle$ silicon handle. The fabrication details of our phononic crystal (PnC) nanoelectromechanical systems (NEMS) resonators are explained elsewhere ~\cite{wollack2021loss,cleland2024studying, Maksymowych2025FreqNoise}. The device we use in this work consists of an array of 9 resonators with slightly different defect cell dimensions, which enables frequency multiplexing. In addition to the resonator highlighted in the main text, we performed similar experiments on the other resonators in the array. Each resonator is found to have its own unique set of TLSs with varying coupling rates.

\section{Measurement Setup}
\begin{figure}[H]
	\centering
	\includegraphics[width=1\textwidth]{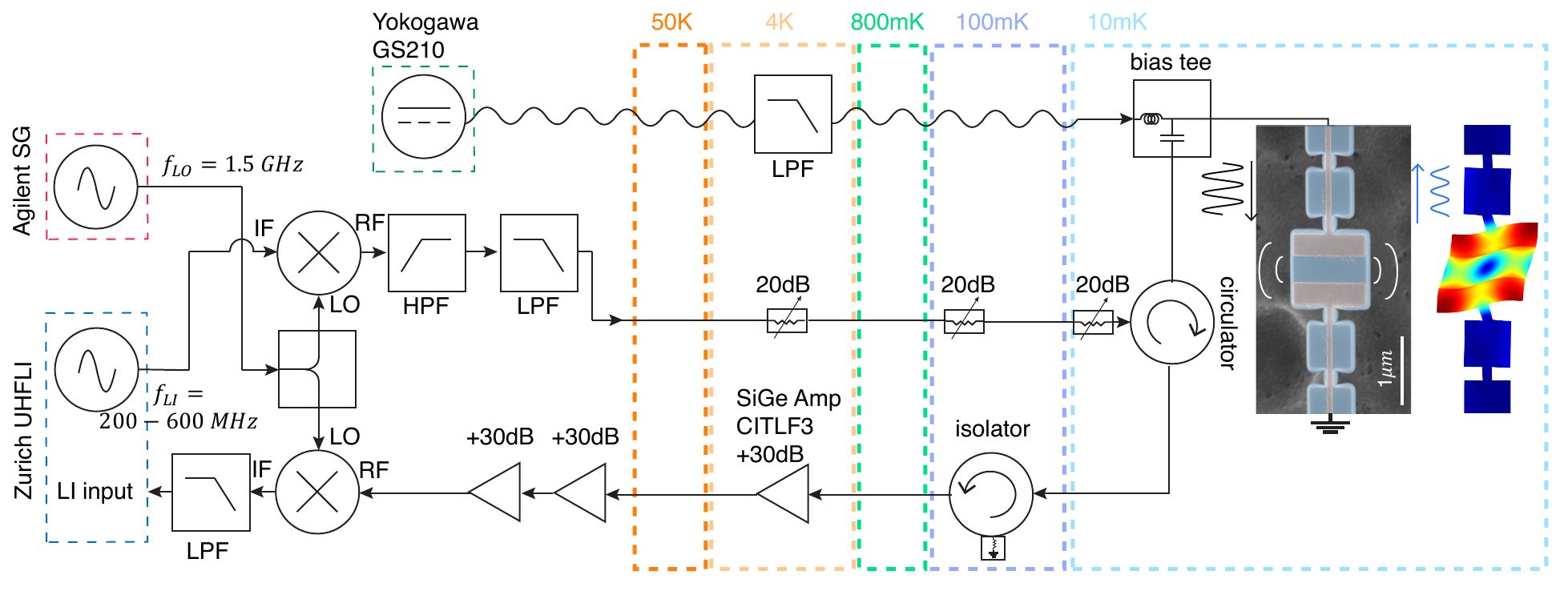} 
	
	\caption{\textbf{Measurement Setup.}
	     Schematic of the horizontal dilution refrigerator measurement setup used to acquire all main text experimental data at Caltech. The microwave drive signal is generated by mixing the local oscillator (LO) tone from a signal generator (SG) at $1.5 \text{ GHz}$ with the Zurich lock-in (LI) output in the DC - 600 MHz range. The signal is up-converted to near the  resonance frequency ($\omega_m/2\pi \approx 1.87 \text{ GHz}$) and the sidebands are removed by cascading the  microwave tone through high pass (HPF) and low pass filters (LPF). The drive signal is then combined with the DC bias at the 10 mK stage via a bias tee circuit. The reflected signal is amplified at $4$ K and at room temperature before being down-converted using the same LO tone and sent to the LI input for demodulation. 
		}
	\label{SI_fig:meas_setup} 
\end{figure}

We perform microwave reflection measurements to investigate the steady-state response of our piezoelectrically actuated mechanical resonators. The full experimental configuration is shown in \ref{SI_fig:meas_setup}. In this approach, the transmitted component of the microwave tone drives the mechanical motion via the piezoelectric effect, while the reflected microwave signal encodes information about the system's response.

For the data presented in the main text, we employ a Zurich Instruments UHFLI lock-in amplifier (LI) to measure on-chip devices thermalized to the mixing chamber plate of a Bluefors horizontal (LH-400) dilution refrigerator (see \ref{SI_fig:meas_setup}). Due to the limited frequency range of the LI (0–600 MHz), the near-resonant drive signal ($\sim 1.9$ GHz) is generated by mixing the output of the LI with a local oscillator (LO) at GHz frequency. After the mixing operation, the undesired sideband is filtered and the signal is transmitted to the device then passed to a superconducting output line via a cryogenic circulator (QCY-G0150201AM). The reflected signal is then mixed down to the LI frequency by the same LO, then filtered before reaching the LI input for demodulation. The LI and LO clocks are frequency locked. The reflected signal is amplified in two stages: first at the $4\text{ K}$ stage using a SiGe cryogenic amplifier (Cosmic Microwave CITLF3), and subsequently at room temperature using additional low-noise amplifiers (Mini-Circuits) as needed. An isolator at the 100 mK stage prevents amplifier input noise from reaching the device. The refrigerator drive line is highly attenuated and constructed from lossy beryllium copper (BeCu) coaxial cables to suppress thermal fluctuations from room temperature down to the $10 \text{ mK}$ stage. Meanwhile, the readout line consists of superconducting niobium titanium (NbTi) coaxial cables from  $10 \text{ mK}$ to $4\text{ K}$ to minimize signal loss. A low-noise DC supply (Yokogawa; GS210) generates the static electrical bias across the device electrodes. The DC signal is routed through twisted-pair cables with sufficient thermalization at each stage within the dilution refrigerator, and a low-pass filter at the 4K stage further reduces the high frequency noise. Just before reaching the device, the microwave drive and DC bias are combined using a bias tee circuit. 

The LI method was utilized at Caltech. Similar measurements were performed on lower frequency phononic crystal devices at Stanford using a vector network analyzer (see section \ref{sec:StanfordMeasurements}). The LI method enables high-sensitivity detection by isolating the signal at a specific reference frequency, improving the signal-to-noise ratio, which is especially advantageous for low-power cryogenic measurements with a low signal-to-noise ratio.

\section{Strong Coupling Modeling and Simulations} \label{sec:StrongCouplingSims}

The simulated spectra of the double avoided crossing of individual TLSs are generated by solving the master equation Equation~\ref{eq:EOM} for each DC detuning while varying the $\omega_\text{TLS} (V_\text{dc})$ according to the relation given in Equation~\ref{eq:TLS}. TLS,C1 and TLS,C2 frequency parameters ($\Delta_0$, $\partial \varepsilon/\partial V_\text{dc}$, $\varepsilon_0$), which determine their dependence on DC bias and yield excellent agreement with the experimental data, are provided in Table~\ref{tab:tls_parameters}. The NEMS-TLS coupling rate $\mathrm{g}$ is obtained by fitting the avoided crossing spectra in Fig.~\ref{fig2}E. For all subsequent simulations, including those investigating power and temperature dependencies, the parameter set $\{\Delta_0,\varepsilon,\partial \varepsilon/\partial V_\text{dc},\mathrm{g}\}$ remains fixed, with only minor adjustments if the TLS frequency shifts slightly.

In addition to these fixed parameters, we allow the following parameters to vary with input drive power and temperature: the resonator's internal decay rate $\kappa_i$, the TLS decay rate $\gamma_\text{TLS}$, and the effective thermal occupancy $n_{eff}$. We also apply small adjustments to the external coupling rate $\kappa_e$ to obtain optimal agreement. We determine these parameters based on best-fit estimates to the experimental data. To fit a single $| S_{11} |$ trace at a chosen DC bias, we first estimate $\kappa_i$ and $\kappa_e$ from the far-detuned response of the resonator - \textit{i.e.} at large DC offset, where the resonator exhibits a single Lorentzian dip largely unaffected by the strongly coupled TLS. To fit to the split spectrum in the presence of strong coupling, we then employ a least-squares optimization to refine $\kappa_i, \kappa_e, \gamma_\text{TLS}, \text{ and } n_{eff}$ by minimizing the mean squared error (MSE) between the simulated and experimental magnitudes of $| S_{11} |$. Concretely, $| S_{11}^{\mathrm{exp}}(\Delta_{m,i})|$ denotes the experimentally measured magnitude at the detuning $\Delta_{m,i}$, and $| S_{11}^{\mathrm{sim}}(\Delta_{m,i}; \mathbf{p})|$ is the simulated magnitude obtained by numerically solving the master equation for a parameter set 
$\textbf{p} = \{\kappa_i, \kappa_e, \gamma_{\text{TLS}}, n_{eff}\}$. We then define:  

\begin{equation}
\label{eq:mse}
\mathrm{MSE}(\mathbf{p}) = \frac{1}{N}\,
\sum_{i=1}^{N}
\left(| S_{11}^{\mathrm{sim}}(\Delta_{m,i};\mathbf{p})| 
   \;-\;
   | S_{11}^{\mathrm{exp}}(\Delta_{m,i}) |
\right)^2,
\end{equation}

\noindent
where $N$ is the total number of detuning points in that sweep. After several iterations, this procedure yields a set of parameters $\mathbf{p}'=\{\kappa_i',\kappa_e',\gamma_{\text{TLS}}', n_{eff}\} = \text{argmin}_\mathbf{p}\text{MSE}(\mathbf{p})$ that provides good agreement with the experimental data, as demonstrated in the main text. Numerically simulating the system at high power and temperature requires a larger Fock-state basis in the master equation, increasing computational complexity. 

The parameter $n_{eff}$ accounts for the increased phonon occupancy arising from higher drive power or elevated temperature. We estimate the uncertainty in $n_{eff}$ via the following approach. We fix all parameters in $\mathbf{p}'$ except $n_{eff}$ and perform a one-dimensional sweep of $n_{eff}$ in the vicinity of its initial best-fit value. We then define a threshold level $1.05\times \text{MSE}(\mathbf{p}')$, corresponding to a $5\%$ increase in MSE compared to the minimum MSE. The range of $n_{eff}$ satisfying $\text{MSE}(n_{eff}) \le 1.05\times \text{MSE}(\mathbf{p}') $ is taken to define the error bar of $n_{eff}$ in the main text. An example of this sweep operation is illustrated in Fig.~\ref{SI_fig:n_MSE}. 

The described method above for a single $| S_{11} |$ trace can be applied in two-dimensional parameter space by taking both $\Delta_{m}$ and DC bias $V_\text{dc}$ as a sweep parameter. To capture the entire parameter space, we extend the MSE definition to include all $(\Delta_{m,i}, V_{dc,j})$ points in the spectra:
\begin{equation}
\label{eq:mse_2D}
\mathrm{MSE}_{\text{2D}}(\textbf{p})
=
\frac{1}{N_{\mathrm{total}}}
\sum_{i=1}^{N_{\Delta}}
\sum_{j=1}^{N_{V}}
\left(
   |S_{11}^{\text{sim}}(\Delta_{m,i}, V_{dc,j}; \textbf{p})|
   -
   |S_{11}^{\text{exp}}(\Delta_{m,i}, V_{dc,j})|
\right)^2,
\end{equation}
\noindent
where $N_{\Delta}$ is the number of detuning points, $N_{V}$ is the number of DC bias points, and $N_{\text{total}} = N_{\Delta} \times N_{V}$ is the total number of points. This approach assumes uncorrelated errors between the simulated and experimental data for each data point of the spectra. By minimizing $\text{MSE}_{\text{2D}}$ over the full $(\Delta_{m}, V_\text{dc})$ plane, we obtain an optimized set of $\{\kappa_i,\kappa_e,\gamma_{\text{TLS}}, n_{eff}\}$ that reproduces the system's response over a DC range that captures the TLS crossings. We find that optimization around critical DC tunings, where $\omega_m \approx \omega_\mathrm{TLS}$, yields parameter values consistent with those from the full 2D optimization. Table~\ref{tab:ps_fits} summarizes all the parameters that are used to produce simulated spectra in Fig.~\ref{fig2} and Fig.~\ref{fig3}, along with the associated fitting errors (MSE, $\text{MSE}_\text{2D}$). For the single-phonon cooperativity values mentioned in the main-text, we use the decay rates of TLS and the mode for $n_{eff} \approx 1$ from the Table~\ref{tab:ps_fits}.

\section{Resolving Random Telegraph Signals}

In our measurements resolving random telegraph signals (RTS), we fix the microwave drive frequency and continuously monitor the device's reflection response $S_{11} (t)$ via the LI, which demodulates the reflected signal to extract its amplitude and phase. The LI's time constant limits the fastest switching process that we can resolve -- a shorter time constant enhances temporal resolution, but it can degrade the signal-to-noise ratio (SNR) as more instrumentation noise is demodulated. Therefore, we carefully optimize the measurement parameters to capture the RTS while maintaining high SNR. For example, the RTS demonstrated in Fig.~\ref{fig5}A was recorded with an LI time constant of  $\sim 4.5~\text{ms}$
at a 5~kHz data acquisition rate over 20 seconds, capturing roughly 80 switching events.

To analyze the data, we calculate the probability distribution of the amplitude response using kernel density estimation (KDE)~\cite{bowman1997applied}, as mentioned in the main text. This method provides a smooth estimate of the underlying distribution, allowing us to resolve closely spaced peaks in the presence of RTS. Once these peaks are identified, each point in the amplitude time trace is assigned to the nearest peak, producing a ladder plot (see red line, Fig.~\ref{fig5}A) that reveals when the system transitions between states.

Once the two-valued ladder plot is constructed, we can easily determine the switching instants and corresponding dwell times—that is, the durations for which the system remains in each state (\textit{i.e.}, classical or quantum). These dwell times are then binned using Scott's rule~\cite{scott1979optimal} to produce a histogram that is normalized to form a probability density function, ensuring that the total area under the histogram equals unity. Fitting the probability distributions to an exponential function indicative of a Poissonian process yields the switching rate  $\lambda$, and the uncertainty in  $\lambda$ is inferred from the $95 \%$ confidence interval of the exponential fit. 

In the presence of RTS, the probability distribution of $|S_{11}|(t)$ exhibits two distinct peaks. These peaks are used to construct the $|S_{11}|$ spectrum in Fig.~\ref{fig5}C. In  Fig.~\ref{SI_fig:RTS_TLSC1}A, we revisit this spectrum and focus on measurements around the resonance frequency. In figure~\ref{SI_fig:RTS_TLSC1}B, we demonstrate how the probability distribution peaks from $|S_{11}|(t)$ evolve around these frequencies. The switching rates from these measurements are demonstrated in Fig.~\ref{fig5}D. Constructing the $|S_{11}|$ spectrum from these peak values provides strikingly clear evidence that the RTS arises from the fluctuations between the split and linear response. 

We observe this RTS behavior in TLS,C1 at elevated temperatures. Figure~\ref{SI_fig:bistability_temperature} shows the same analysis performed over a temperature range from 400~mK to 800~mK, all showing split and linear responses evident. Remarkably, we find evidence of strong coupling even at significantly high temperatures, as high as 800~mK. 

\section{Estimating the TLS Spectral Density and the Phonon-TLS Coupling Strength}
\label{sec:EstimatingCoupling}
Previous effects observed below 1~K in phononic crystals, such as frequency redshifts and quality factor tuning~\cite{wollack2021loss,cleland2024studying, emser2024thin}, have been explained using the standard tunneling model (STM)~\cite{phillips1972tunneling,anderson1972anomalous,phillips1987two}. This model assumes a large number of solid-state TLS in the material, which have tunneling and asymmetry energies distributed over a wide interval. This suggests that a continuum of TLS are expected to interact with our mode. The STM also predicts weak TLS-mode and TLS-TLS coupling. In contrast, we measure TLSs that are discretely distributed, strongly coupled to the mode, and may be coupled to each other leading to telegraph noise. We have also reported in a separate manuscript that individual strongly coupled TLS and TLS-TLS coupling may be critical for explaining telegraph noise even at high drive powers ~\cite{Maksymowych2025FreqNoise}. All of these observations may be understood by taking into account the small size of the resonators we study. To count the number of TLS involved in mode interactions we consider the structurally disordered and damaged regions of the device, which all reside near the surface. Assuming a 5~nm thick aluminum oxide, 5~nm thick aluminum-LN interface, a 5~nm LN-Si bonding layer, and a 5~nm thick damaged LN layer (from ion milling~\cite{gruenke2024surface, gruenke2025surface}), we calculate a TLS host volume of $V_h\approx 0.046 \text{ }\mu \text{m}^3$ given our PnC resonator geometry and dimensions. With an estimate of the TLS density for disordered solids $P_0\sim 10^{44}$/Jm$^3$ \cite{black1978relationship, behunin2016dimensional,chen2024phonon} we find the frequency density (spacing) of TLS to be $\rho_{\text{TLS}}=P_0V_h\approx 3~\text{TLS/GHz}$, which is highly discontinuous. Similar calculations were reported in ~\cite{Maksymowych2025FreqNoise, bozkurt2024mechanical}.

Furthermore, we may use the TLS elastic dipole moment $\gamma$ (deformation potential) to estimate the TLS-phonon coupling strength, $\mathrm{g}\approx |\gamma|\xi_{zp}/\hbar$ ~\cite{Maksymowych2025FreqNoise}, where $\xi_{zp}=\sqrt{\hbar\omega_m/2\mathcal{E}V_m}\approx 6.5\times10^{-9}$m/m is the RMS zero-point strain of the mechanical mode, $V_m = \int_V\xi^{\mu\nu}c_{\mu\nu\alpha\beta}\xi^{\alpha\beta}d^3\mathbf{r}/\text{max}(\xi^{\mu\nu}c_{\mu\nu\alpha\beta}\xi^{\alpha\beta}) \approx 0.16\text{ } \mu \text{m}^3$ is the acoustic mode volume simulated via COMSOL, $\mathcal{E}$ is the elastic modulus of LN, and $c_{\mu\nu\alpha\beta}$ is the  stiffness tensor of LN. Recalling that $\gamma\equiv(1/2)\partial\varepsilon/\partial\xi = (1/2)\times\partial\varepsilon/\partial V \times \partial V/\partial\xi$, we can simulate the volumetric LN crystal strain induced by DC biasing in COMSOL, $\partial\xi/\partial V\approx$ 7.5$\times10^{-5}$/V and we know $\partial\varepsilon/\partial V\approx$ $h\times$30~GHz/V from direct TLS measurements. Therefore, we estimate $|\gamma|\approx$ 0.84~eV and $\mathrm{g}/2\pi\approx $ 1.5~MHz, roughly equivalent to our experimental results of 1.7~MHz and 2.4~MHz. The inferred dipole moment of TLS in LN is comparable to the TLS present in silica \cite{golding1976phonon,maccabe2020nano,bozkurt2024mechanical,bozkurt2023quantum}. 

\section{Additional Experimental Data}
\label{sec:additional_data}
We performed a comprehensive range of experiments to understand the interactions between TLS defects and the mechanical mode that support and build upon the findings in the main text. The remarkable stability and reproducibility of the TLS frequency, loss, and phonon-TLS coupling strength of TLS,C1 enabled a multitude of experiments. Figure~\ref{SI_fig:TLSC1_fine}B shows a DC bias sweep taken with a very fine voltage stepsize of $50~\mathrm{\mu V}$ near the single avoided crossing at different incident microwave drive powers $P$. This is much finer than Fig.~\ref{SI_fig:TLSC1_fine}A and the other spectra in the main text, which were taken with $500\, \mathrm{\mu V}$ steps. We consistently excite higher or lower energy-level transitions by adjusting the input power, demonstrating a means to control the most favored energy transitions of the coupled system.  

\begin{figure}[H] 
	\centering
	\includegraphics[width=0.8\textwidth]{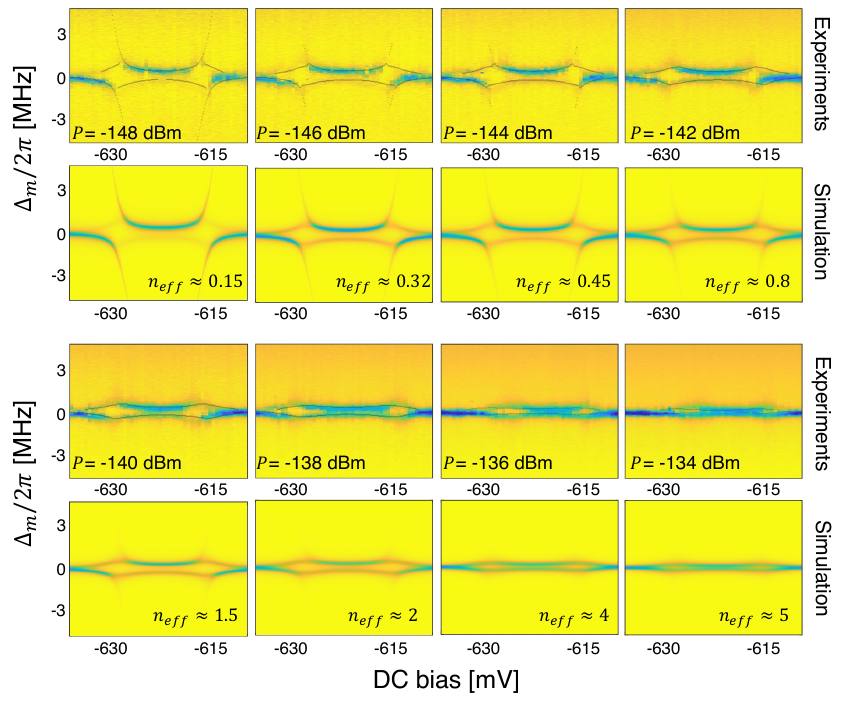} 
	
	\caption{\textbf{Power Saturation of TLS,C2.}
	     Power sweep measurements performed with TLS,C2 demonstrated in Fig.~\ref{fig2}E. The plot includes experimental data and simulated spectra. The minimas of simulated $| S_{11} |$ are overlaid on the experimental data as black lines to demonstrate agreement with simulations. Compared to TLS,C1, we observe slower increase in $n_{eff}$ with power, possibly due to the lower phonon-TLS coupling strength compared to TLS,C1.
		}
	\label{SI_fig:TLSC2} 
\end{figure}

\begin{figure}[H] 
	\centering
	\includegraphics[width=1\textwidth]{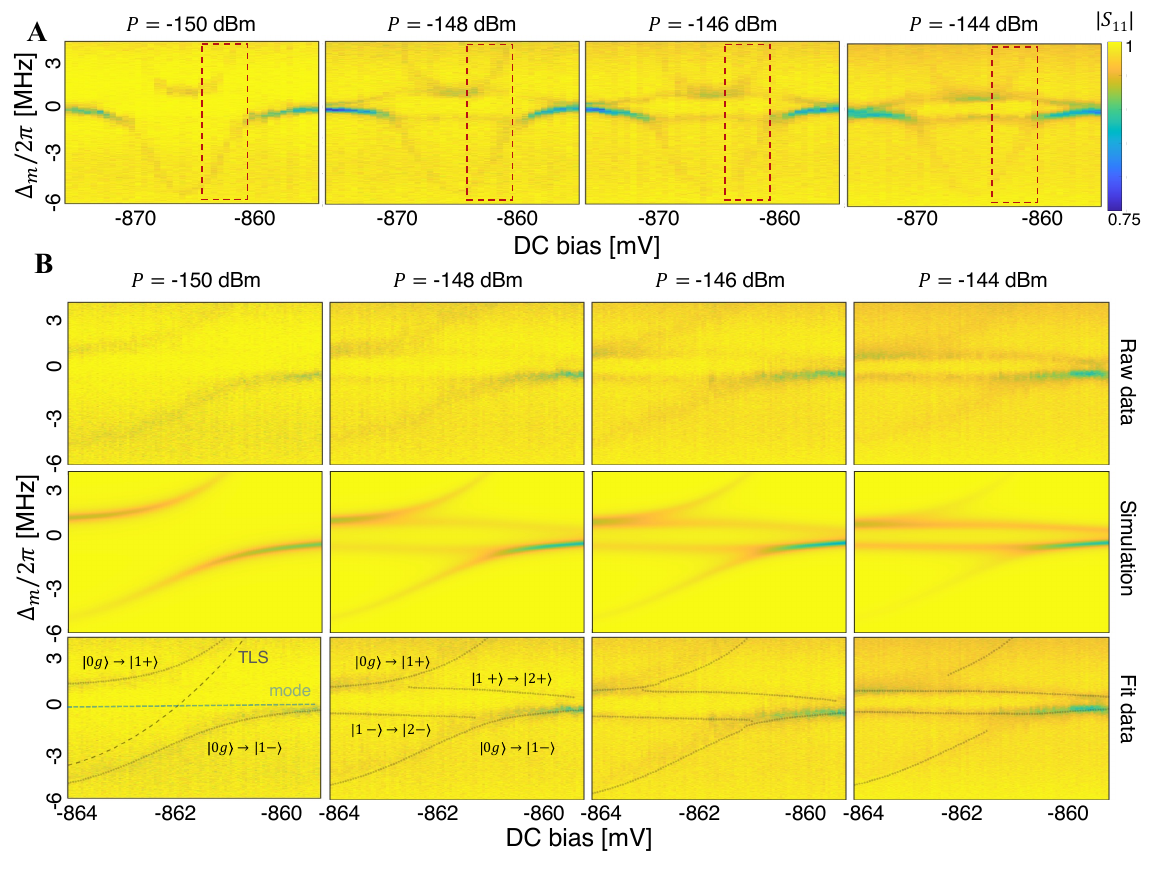} 
	
	\caption{\textbf{Finer Measurements of JC Ladder Features.}
	    (\textbf{A}) The red square highlights the region where finer sweeps were performed around one of the avoided crossings on TLS,C1. (\textbf{B}) A DC sweep in 50 $\mu V$ increments reveals the fine structure of the crossing and the higher energy transitions as we increase the input power. The stability of the NEMS-TLS system enables continuous and repeatable measurements in the strong coupling regime. Excellent agreement is observed between the experimental results and simulations.
		}
    \label{SI_fig:TLSC1_fine} 
\end{figure}

Although we mostly focused on TLS,C1 in the main text due to its high coupling rate $\mathrm{g}$ and tunneling rate $\Delta_0$ being near mechanical mode frequency $\omega_m$, we also perform the same power sweep experiments with TLS,C2 by tuning the bias to near $\sim-630$~mV. Much like TLS,C1, Fig.~\ref{SI_fig:TLSC2} shows that increasing the incident power induces higher-level transitions, giving rise to an \textit{eye-like} crossing. The extracted $n_{eff}$ values indicate a weaker dependence on power for TLS,C2 compared to TLS,C1, likely due to the smaller phonon-TLS coupling strength of 1.7~MHz, compared to 2.4~MHz for TLS,C1. 

\begin{figure}[H] 
	\centering
	\includegraphics[width=1\textwidth]{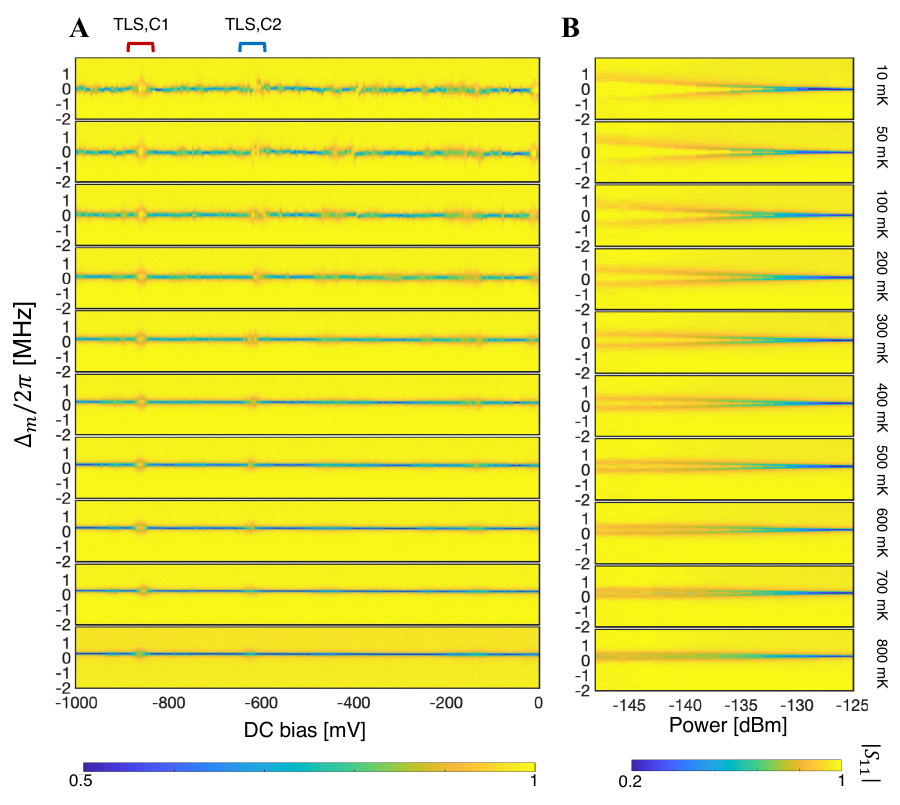} 
	
	\caption{\textbf{Probing the Discrete TLS bath at Different Stage Temperatures.}
	    (\textbf{A}) A wider DC sweep from -1 to 0 V reveals a variety of spectral features at lower temperatures, most of which saturate at higher temperatures. TLS,C1, the strongest coupled TLS we found, maintains its characteristics even at 800 mK.  (\textbf{B}) Transition to the classical response of TLS,C1 at different temperatures. This figure complements Fig.\ref{fig4}B,D,F by presenting the full temperature sweep. As the temperature increases, the splitting is less sensitive to the input power as we observe constant splitting over a wider power range. 
		}
	\label{SI_fig:wideDC_T} 
\end{figure}

\begin{figure}[H]
    \centering
    \includegraphics[width=0.5\linewidth]{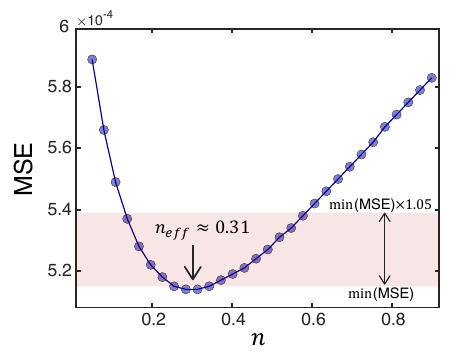}
    \caption{\textbf{Characterization of $5\%$ Tolerance Criterion for $n_{eff}$.}
    The plot displays the mean squared error (MSE) as a function of the estimated phonon number $n$. After iterative minimization of the fitting parameters, we sweep $n$ around the optimized $n_{eff}$ value to quantify the uncertainty in our estimation. This sweep identifies the regions $n<n_{eff}$ and $n>n_{eff}$ where the MSE increases by $5\%$ above its minimum value (indicated by the shaded area). Notably, the error rises more steeply for $n<n_{eff}$ than for $n>n_{eff}$, as expected due to the $\sqrt{n}$ scaling of the energy levels. As $n_{eff}$ increases further, achieving an accurate determination becomes increasingly challenging and computationally expensive.  
    }
    \label{SI_fig:n_MSE}
\end{figure}

\begin{figure}[H]
	\centering
	\includegraphics[width=0.9\textwidth]{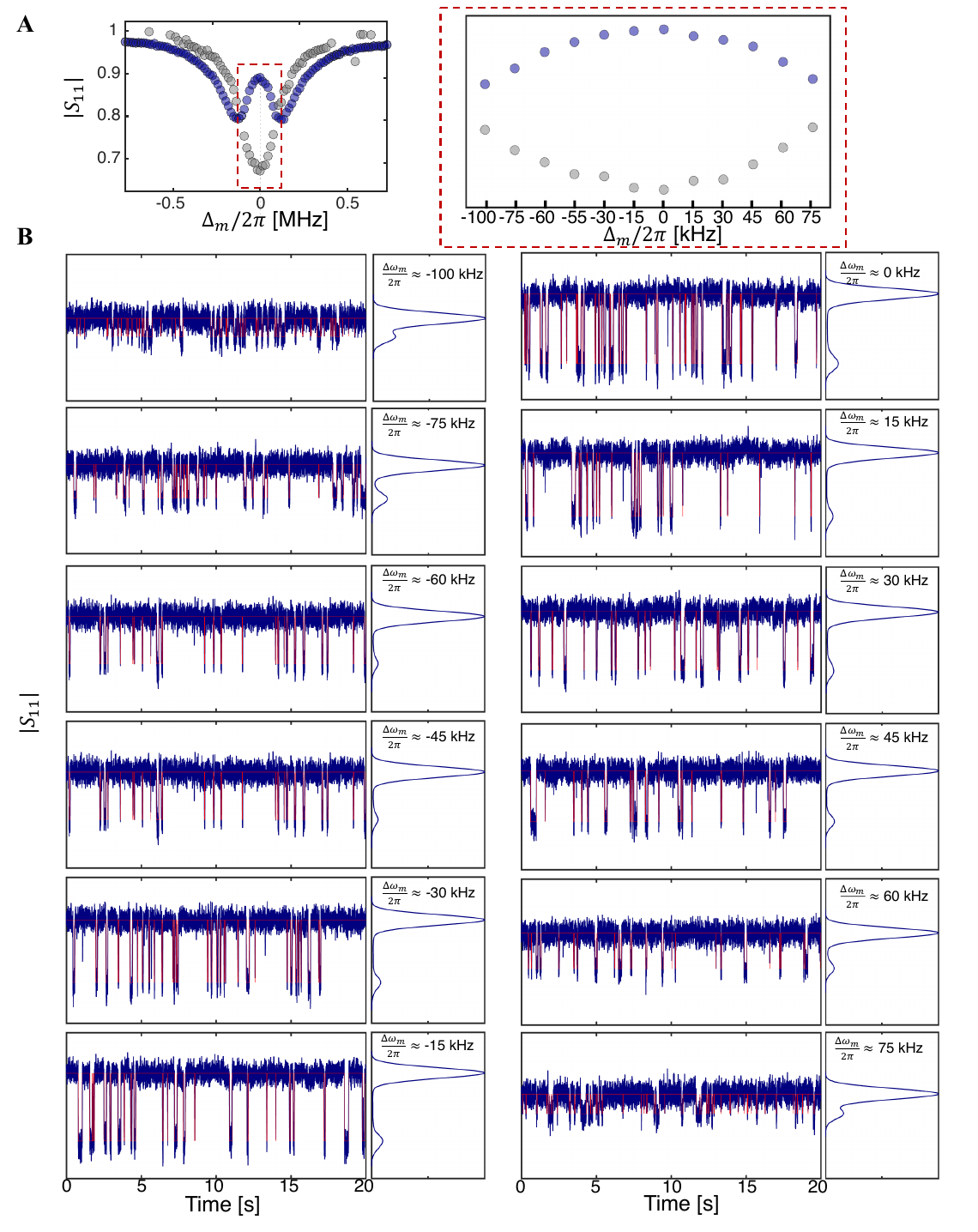} 
	
	\caption{\textbf{RTS Detuning Sweep Analysis.}
	    (\textbf{A}) Time series data used to plot the data points in the red square. On the right, we zoom into this region and provide the detuning values on the x-axis. (\textbf{B}) Data is collected over 20 seconds, and the corresponding probability density plot is provided on the right for each detuning within the same $|S_{11}|$ interval. The distribution reveals fluctuations between two distinct peaks, which are used to plot the reflection coefficient, $|S_{11}|$. 
		}
	\label{SI_fig:RTS_TLSC1} 
\end{figure}

\begin{figure}[H] 
	\centering
	\includegraphics[width=1\textwidth]{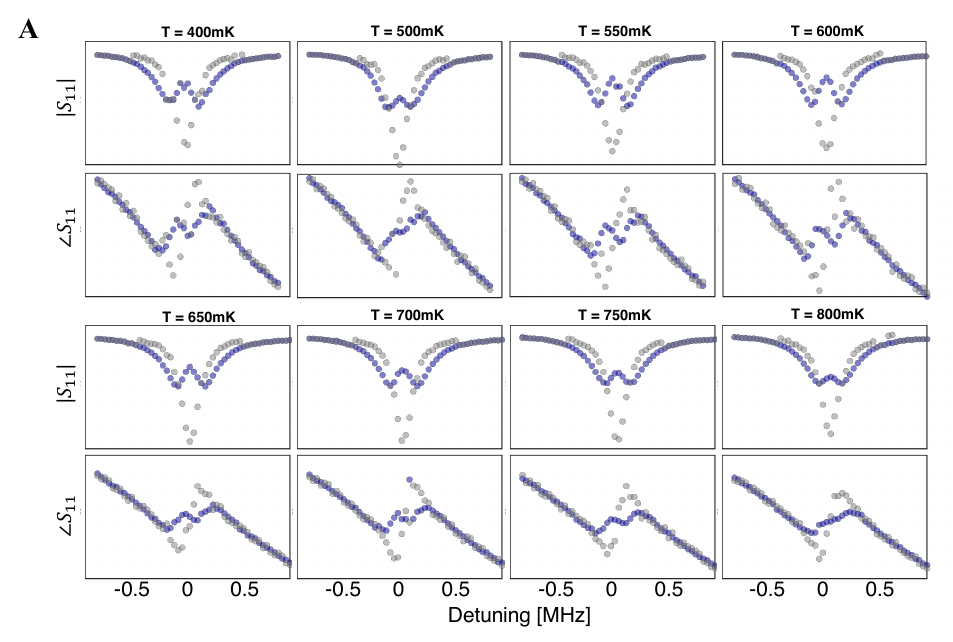} 
	
	\caption{\textbf{RTS Behavior versus Stage Temperatures.}
	    (\textbf{A}) Measurements similar to those demonstrated in Fig.\ref{fig5} and \ref{SI_fig:RTS_TLSC1} are repeated for varying temperatures. RTS behavior is very stable which allows us to perform more statistical analysis on the data.
		}
	\label{SI_fig:bistability_temperature} 
\end{figure}

Figure~\ref{SI_fig:wideDC_T} illustrates the evolution of TLS crossings over a wide range of DC as the temperature increases, while the incident power remains fixed at $-140~\text{dBm}$. We observe more crossings at lower temperatures, as at high temperatures most TLSs saturate. The most strongly coupled defects (TLS,C1 and TLS,C2) consistently appear at the same DC bias voltage despite us repeatedly sweeping the DC bias over a broad range and changing the temperature for each measurement, from $10 - 800~\text{mK}$. This highlights the stability of the strongly coupled TLS.

\begin{figure}[H]
	\centering
	\includegraphics[width=0.9\textwidth]{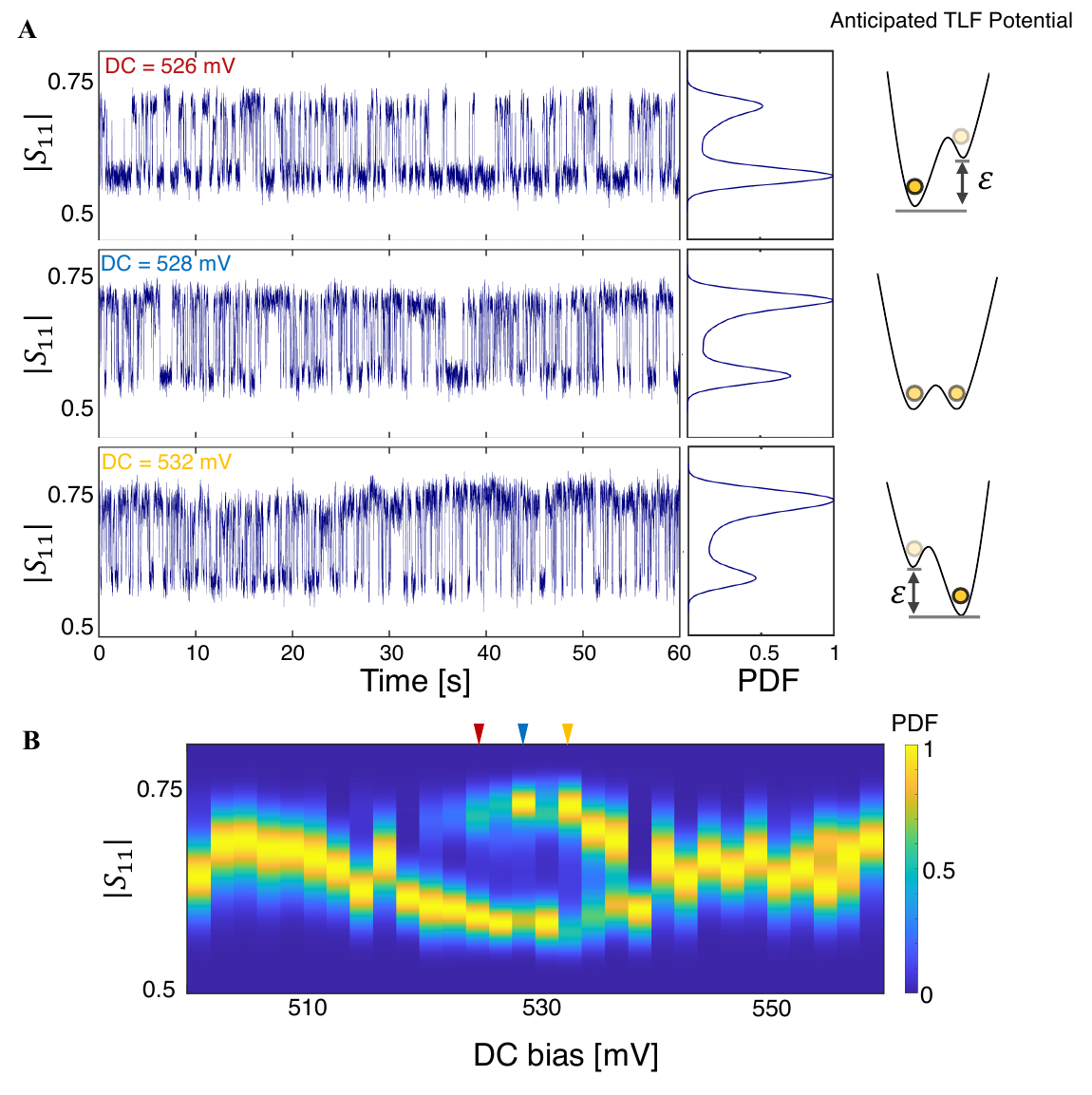} 
	
	\caption{\textbf{Observation of Tunable RTS Probability Distribution.}
	    (\textbf{A}) RTS in the reflection amplitude $|S_{11}|$ acquired at $50 \text{mK}$ during a different cooldown and using a different resonator from the main text ($\omega_m/2\pi = 1.95 \text{GHz}$). The RTS in the mechanics is observed as the frequency of the strongly coupled TLS hops between on- and off-resonance. The symmetry of the RTS changes with DC bias, which could be explained by changes in the most probable state in the double-well potential of a TLF strongly coupled to the TLS. (\textbf{B}) Heatmap of the normalized probability distributions of the RTS as a function of applied DC bias. The three colored triangles at the top indicate the DC values used in the time series plot in (A). The opening and closing behavior is due to the strongly coupled TLS crossing the mechanics, while changes in the most probable state are likely caused by a fluctuator. 
		}
	\label{SI_fig:RTS_R5} 
\end{figure}

Throughout our study, we could detect RTS at multiple DC voltages across various resonators. In Fig.~\ref{SI_fig:RTS_R5}, we demonstrate an example of RTS observed in different resonator ($\omega_m/2\pi = 1.95~\text{GHz}$) from the main text while at a base plate temperature of $50~\text{mK}$. In this example, we demonstrate that the probability of finding the resonator in one of two stable states can be tuned by DC biasing. This phenomenon could be explained by strong TLS-TLF coupling. Whenever the TLF tunnels between its eigenstates, which are primarily localized in one of the two well minima, the TLS frequency would jump from on- to off-resonance (or vice versa). Here, we show direct control over the double-well potential of the TLF asymmetry and can alter which well the state localizes in (shown schematically on the right). Figure~\ref{SI_fig:RTS_R5}B plots the normalized probability distribution of these RTS events as a function of DC bias. If we increase the DC bias sufficiently far from 530~mV (where the RTS is most symmetric and the TLS is resonant with the mode) we can remove the RTS. This measurement shows that DC biasing may be an effective way of engineering the quantum noise present in the single-phonon regime and might provide an avenue towards noise mitigation. 
\\

\begin{table}[H]
    \setlength{\tabcolsep}{12pt}
    \label{tab:tls_parameters}
    \centering
    
    \begin{tabular}{lcccc}
        \toprule
        \textbf{TLS ID} & $\mathrm{\Delta_0}/h$ \text{[GHz]} & $\mathrm{\varepsilon_0}/h$ \text{[GHz]} & $\frac{1}{h}\mathrm{\frac{\partial \varepsilon}{\partial V_\text{dc}}}$ \text{[GHz/V]} & $\mathrm{g}/2\pi \text{[MHz]}$ \\[1mm]
        \midrule
        TLS,C1 & 1.8692 & 33.26   & 38.42 & 2.40 \\[2mm]
        
        TLS,C2 & 1.8645 & 18.66   & 30.01 & 1.72 \\
        \bottomrule
    \end{tabular}
    \caption{\textbf{TLS Parameters}. This table summarizes the key parameters of TLS,C1 and TLS,C2 used in simulations, where $h$ is Planck's constant.}
\end{table}

\begin{table}[H]
\setlength{\tabcolsep}{16pt}
\resizebox{\textwidth}{!}{
\begin{tabular}{lcccccc}
\toprule
\textbf{$P$ [dBm]} & \text{$\gamma_{\text{TLS}}/2\pi$ [kHz]} & \text{$\kappa_i/2\pi$ [kHz]} & \text{$\kappa_e/2\pi$ [kHz]} & \text{$n_{eff}$} & \text{MSE} & \text{MSE$_\text{2D}$} \\[1mm]
\midrule
    $-150$ & $200$ & $400$ & $17$ & $0.08$ & $5.90 \times 10^{-4}$ & $2.12 \times 10^{-4}$  \\[2mm]
    $-149$ & $184$ & $180$ & $18$ & $0.22$ & $7.80 \times 10^{-4}$ & $3.55 \times 10^{-4}$\\[2mm]
    $-148$ & $169$ & $160$ & $22$ & $0.29$ & $5.14 \times 10^{-4}$ &$2.49 \times 10^{-4}$ \\[2mm]
    $-147$ & $150$ & $150$ & $25$ & $0.31$ & $5.36 \times 10^{-4}$ & $2.57 \times 10^{-4}$\\[2mm]
    $-146$ & $106$ & $100$ & $24$ & $0.41$ & $4.34 \times 10^{-4}$ &$ 2.49 \times 10^{-4}$  \\[2mm]
     $-145$ & $81$ & $112$ & $17$ & $0.40$ & $2.88 \times 10^{-4}$ &$1.94 \times 10^{-4}$ \\[2mm]
     $-144$ & $50$ & $100$ & $19$ & $0.68$ & $1.99 \times 10^{-4}$ & $1.56 \times 10^{-4}$ \\[2mm]
    $-143$ & $87$ & $95$ & $20$ & $0.73$ & $1.52 \times 10^{-4}$ & $1.26 \times 10^{-4}$\\[2mm]
    $-142$ & $65$ & $100$ & $24$ & $0.99$ & $1.58 \times 10^{-4}$ & $1.22 \times 10^{-4}$\\[2mm]
    $-141$ & $55$ & $80$ & $24$ & $1.37$ & $0.99 \times 10^{-4}$ & $0.99 \times 10^{-4}$ \\[2mm]
    $-140$ & $84$ & $80$ & $29$ & $1.50$ & $0.90 \times 10^{-4}$ & $1.01 \times 10^{-4}$\\[2mm]
    $-139$ & $60$ & $67$ & $29$ & $1.75$ & $0.77 \times 10^{-4}$ &$1.05 \times 10^{-4}$ \\[2mm]
    $-138$ & $25$ & $84$ & $30$ & $3.57$ & $0.69 \times 10^{-3}$ & $0.78 \times 10^{-4}$ \\[2mm]
    $-137$ & $23$ & $71$ & $30$ & $3.84$ & $0.54 \times 10^{-4}$ &$0.78 \times 10^{-4}$  \\[2mm]
    $-136$ & $30$ & $86$ & $35$ & $5.27$ & $0.77 \times 10^{-4}$ &$0.89 \times 10^{-4}$  \\[2mm]
\bottomrule
    \end{tabular}}
\caption{\textbf{Optimized Parameters for TLS,C1 Simulations}. The listed parameters were used to simulate the master equation and achieve a close match to the experimental data shown in  Fig.~\ref{fig3}. MSE values represent the error calculated for fixed DC bias in Fig.\ref{fig3}C, while $\text{MSE}_{2\text{D}}$ corresponds to the errors associated with the spectra in Fig.\ref{fig3}B. }
\label{tab:ps_fits}
\end{table}

\newpage
\section{Strong Coupling between Individual TLS and Sub-Gigahertz Phononic Crystal Resonators} \label{sec:StanfordMeasurements}

\begin{figure}[H]
    \centering
    \includegraphics[width=1\linewidth]{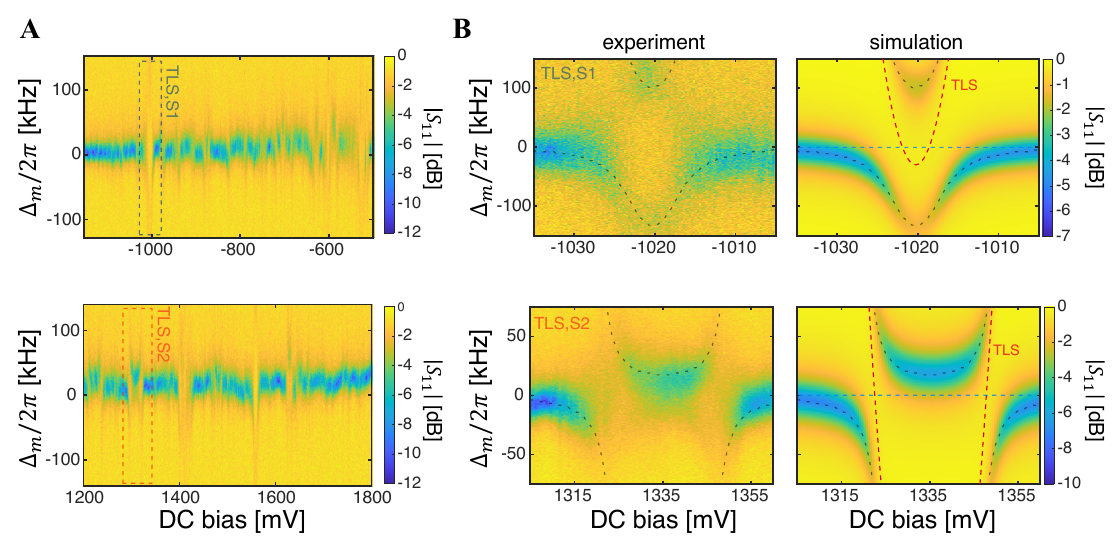}
    \caption{\textbf{Probing the Discrete TLS Bath and the Double Avoided Crossings of TLS,S1 and TLS,S2.}
    Panels in (\textbf{A}) show the magnitude of the reflection S-parameter $|S_{11}|$ of the phononic crystal resonator $(\omega_m / 2\pi \approx751 \text{ MHz})$ detailed in Supplementary Table \ref{tab:tls_parameters_stanford} as a function of DC bias. While many double avoided anti-crossings are visible, the ones corresponding to TLS,S1 and TLS,S2 that we studied in detail are outlined with dashed boxes. For this measurement we employed a continuous coherent drive of -156 dBm with the device stage at 10 mK, ensuring that the phonon occupancy near-zero ($n_{th}\approx 0.03$). (\textbf{B}) Finer DC sweeps around TLS,S1 and TLS,S2 with the simulated spectra with $P$ = -164 dBm. The corresponding coupling rates $\mathrm{g}/2\pi$ are determined to be 120 kHz and 85 kHz, respectively.
    }
    \label{SI_fig:Stanford_TLS}
\end{figure}

\begin{figure}[H]
    \centering
    \includegraphics[width=0.7\linewidth]{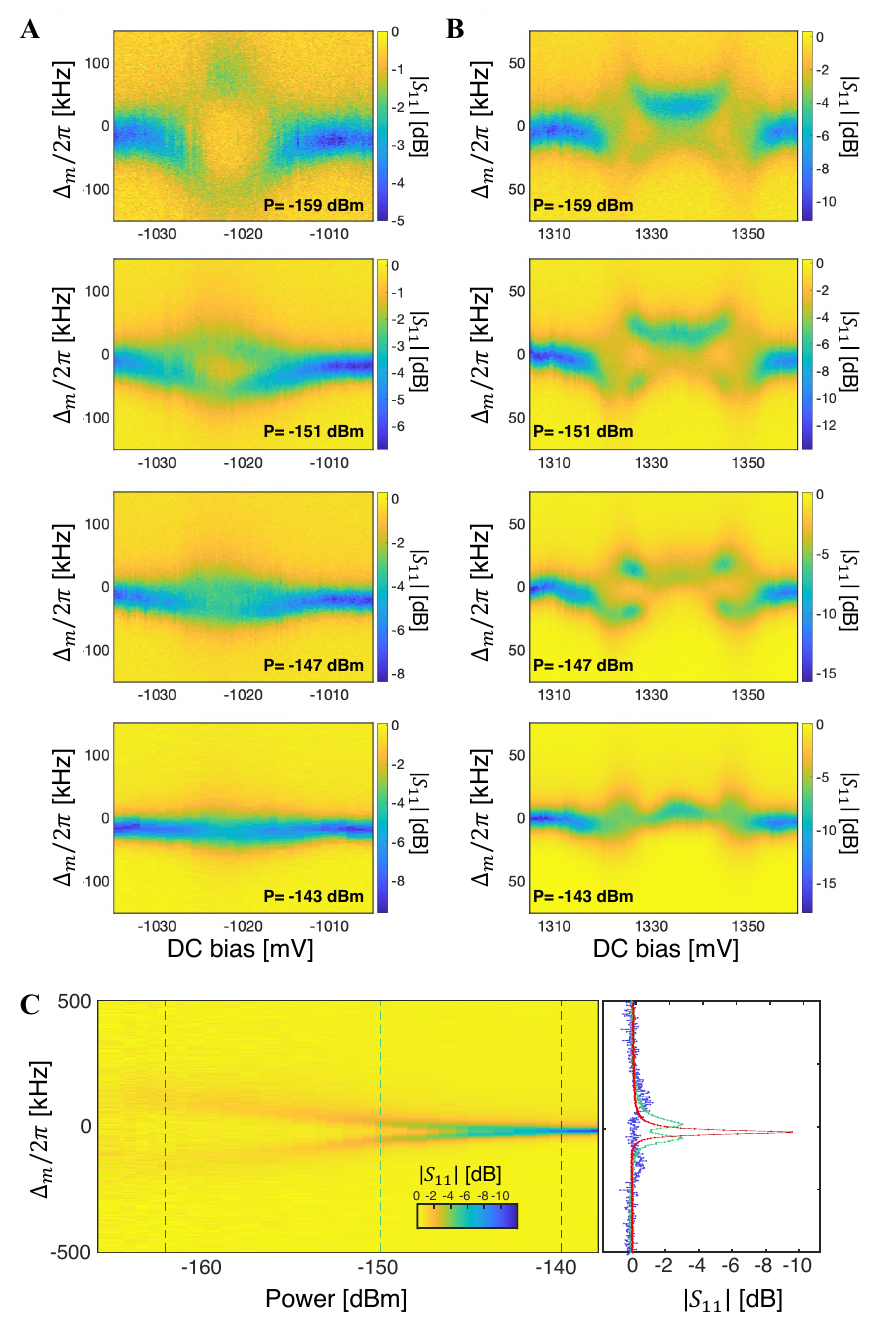}
    \caption{\textbf{Observation of Power Saturation of TLS,S1 and TLS,S2.}
    The reflection parameter $|S_{11}|$ versus DC bias is shown for TLS,S1 in (\textbf{A}) and TLS,S2 in (\textbf{B}) at different incident microwave powers $P$. As the incident power increases, we observe the higher energy level transitions followed by a transition to the classical response. (\textbf{C}) We show the quantum to classical transition by measuring the reflection S-parameter $|S_{11}|$ spectrum as a function of drive power while at a fixed DC bias of -1022 mV for TLS,S1. On the right panel, we plot the individual spectra of $|S_{11}|$ measured with a $P$ of -164 dBm, -150 dBm and -140 dBm as purple, green and red, respectively - marked with vertical dashed lines in the heat map. 
    }
    \label{SI_fig:StanfordPowerSweep}
\end{figure}

We measured similar strong coupling phenomena using devices from a separate nanofabrication run measured at Stanford. These devices are high-quality phononic crystal resonators with fundamental shear mode frequencies near 750 MHz ~\cite{Maksymowych2025FreqNoise}. The measurements we discuss here were performed at 10 mK in a vertical dilution refrigerator (Bluefors; LD250) using the setup detailed in~\cite{Maksymowych2025FreqNoise}. 
As in the main text, we apply a tunable DC bias across the PCR electrodes to shift the asymmetry energy (i.e. frequency) of individual TLS with respect to the relatively unchanging mechanical mode frequency. We measure the mode by microwave reflectometry using a vector network analyzer (VNA), which gives the S-parameter, $S_{11}$. In \ref{SI_fig:Stanford_TLS}A we show $|S_{11}|$ (dB) as a function of frequency detuning from the mechanical resonance $\Delta_m/2\pi$ over a broad range of DC bias voltages taken with a coherent drive of -156 dBm ($\bar{n} = P/\hbar\omega_m \times 4\kappa_e/\kappa^2\approx 0.66$) at 10 mK ($n_{th}\approx 0.03$). We observe multiple double avoided anti-crossings at distinct values of DC bias. Like the main text, we attribute these splittings to individual TLS defects which strongly hybridize with the mechanical resonator when tuned onto resonance. On-plot we have indicated two regions by dashed boxes where individual TLS cross the mechanics. In \ref{SI_fig:Stanford_TLS}B, we show data with a finer DC voltage stepsize (250 $\mu$V) taken near these two double anti-crossings, each of which is associated with one TLS (TLS,S1 and TLS,S2). Dashed red lines indicate the TLS hyperbolic trajectories described by Eq.~\ref{eq:TLS}. We can model the data using the approach from the main text and \ref{sec:StrongCouplingSims}. The parameters of the two TLSs are in Supplementary Table \ref{tab:tls_parameters_stanford}. In \ref{SI_fig:StanfordPowerSweep}A and B, we see direct evidence of power saturation of these TLS resulting in an \textit{eye-like} crossing as predicted by the JC model. Next, we park the DC bias at  $-1022$ mV near TLS,S1, then ramp the RF drive power (see \ref{SI_fig:StanfordPowerSweep}C). At low power, we observe vacuum Rabi splitting characterized by a coupling strength of $\mathrm{g}/2\pi\approx$ 120 kHz. This result tracks well with our calculations in ~\cite{Maksymowych2025FreqNoise}, where we estimated a maximum coupling rate of $\sim$360 kHz. At higher powers, we see the expected transition from the quantum (nonlinear) to classical (linear) resonator behavior, described by a single lorentzian.

The phonon-TLS coupling in this device is lower than for the 1.9 GHz mode in the main text because the higher frequency device has a smaller volume and thus a larger zero-point strain. When detuned far from resonance, these same TLS can dominate the frequency noise at high drive power where resonant TLS are saturated~\cite{Maksymowych2025FreqNoise}. Much like the 1.9 GHz device, we observe a frequency spectral density of TLS that is discrete, not the quasicontinuum predicted by the standard tunneling model. This also agrees with ~\cite{Maksymowych2025FreqNoise}, where we predicted the presence of a discrete TLS bath ($\rho_{TLS}\approx$1 TLS/150 MHz). We note that the reflection spectroscopy was performed with many averages. For instance, for \ref{SI_fig:Stanford_TLS}B we employ a 10 Hz integration bandwidth and 30 averages for each frequency domain scan. This implies that the inferred $\kappa_i$ values incorporate spectral broadening from dephasing and energy decay ~\cite{Maksymowych2025FreqNoise}.
\\

\begin{table}[H]
    \setlength{\tabcolsep}{7pt}
    
    \centering
    
    \begin{tabular}{lccccccc}
        \toprule
        \textbf{TLS ID} & $\mathrm{\Delta_0}/ h$ \text{[MHz]} & $\mathrm{\varepsilon_0}/ h$ \text{[GHz]} & $\frac{1}{h}\mathrm{\frac{\partial \varepsilon}{\partial V_\text{dc}}}$ \text{[GHz/V]} & $\mathrm{g}/2\pi \text{[kHz]}$ & {$\kappa_e/2\pi \text{ [kHz]}$ } & {$\kappa_i/2\pi\text{ [kHz]}$ } & $\gamma_{\text{TLS}}/2\pi \text{ [kHz]}$ \\[1mm]
        \midrule
        TLS,S1 & 751.24 & 4.13   & 4.04 & 120 & 11.1 & 30.2 & 40  \\[2mm]
        TLS,S2 & 750.92 & -5.4   & 1.82 & 85 & 11.5 &40.8 & 60 \\
        \bottomrule
        
    \end{tabular}
    \caption{\textbf{TLS and Phononic Crystal Resonator Parameters}. This table summarizes the key parameters of TLS,S1 and TLS,S2 used for the simulations shown in \ref{SI_fig:Stanford_TLS}B, where $h$ is Planck's constant.}
    \label{tab:tls_parameters_stanford}
\end{table}

\end{document}